\documentclass[a4paper,11pt]{article}
\pdfoutput=1 

\usepackage{jcappub} 

\usepackage[T1]{fontenc} 
\usepackage[numbers]{natbib}
\usepackage{cleveref}

\usepackage{subfig}

\usepackage{siunitx}
\DeclareSIUnit\Mpc{Mpc}
\DeclareSIUnit\Gpc{Gpc}
\DeclareSIUnit\kpc{kpc}
\DeclareSIUnit\Gyr{Gyr}
\DeclareSIUnit\yrs{yrs}
\DeclareSIUnit\yr{yr}

\DeclareSIUnit{\hub}{\mathit{h}}
\DeclareSIUnit{\invhub}{\mathit{h}^{-1}}
\newcommand{\SIF}[2]{\SI[parse-numbers=false]{#1}{#2}}

\usepackage{bm}

\usepackage{tikz}
\usepackage{pgfplots}
\usepackage{pgfplotstable}
\usetikzlibrary{external}
\usetikzlibrary{shapes}
\usetikzlibrary{decorations.text}
\usepgfplotslibrary{fillbetween}
\title{\boldmath Where shadows lie: reconstruction of anisotropies in the neutrino sky}

\author[a]{Willem Elbers,}
\author[a]{Carlos S. Frenk,}
\author[a]{Adrian Jenkins,}
\author[a]{Baojiu Li,}
\author[b,c]{Silvia Pascoli,}
\author[d]{Jens Jasche,}
\author[e]{Guilhem Lavaux,}
\author[f]{and Volker Springel}

\affiliation[a]{Institute for Computational Cosmology, Department of Physics, Durham University, South Road, Durham, DH1 3LE, UK}
\affiliation[b]{Dipartimento di Fisica e Astronomia, Universit\`a di Bologna, via Irnerio 46, 40126 Bologna, Italy}
\affiliation[c]{INFN, Sezione di Bologna, viale Berti Pichat 6/2, 40127 Bologna, Italy}
\affiliation[d]{The Oskar Klein Centre for Cosmoparticle Physics, Department of Physics, Stockholm University, AlbaNova, Stockholm SE-106 91, Sweden}
\affiliation[e]{Sorbonne Université, CNRS, UMR 7095, Institut d’Astrophysique de Paris, 75014 Paris, France}
\affiliation[f]{Max Planck Institute for Astrophysics, Karl-Schwarzschild-Str 1, D-85741 Garching, Germany}

\emailAdd{willem.h.elbers@durham.ac.uk}

\abstract{The Cosmic Neutrino Background (CNB) encodes a wealth of information, but has not yet been observed directly. To determine the prospects of detection and to study its information content, we reconstruct the phase-space distribution of local relic neutrinos from the three-dimensional distribution of matter within $\SIF{200\invhub}{\Mpc}$ of the Milky Way. Our analysis relies on constrained realization simulations and forward modelling of the \texttt{2M++} galaxy catalogue. We find that the angular distribution of neutrinos is anti-correlated with the projected matter density, due to the capture and deflection of neutrinos by massive structures along the line of sight. Of relevance to tritium capture experiments, we find that the gravitational clustering effect of the large-scale structure on the local number density of neutrinos is more important than that of the Milky Way for neutrino masses less than $\SI{0.1}{\eV}$. Nevertheless, we predict that the density of relic neutrinos is close to the cosmic average, with a suppression or enhancement over the mean of $(-0.3\%,\,+7\%,\,+27\%)$ for masses of $(0.01,\,0.05,\,0.1)\,\si{\eV}$. This implies no more than a marginal increase in the event rate for tritium capture experiments like PTOLEMY. We also predict that the CNB and CMB rest frames coincide for $\SI{0.01}{\eV}$ neutrinos, but that neutrino velocities are significantly perturbed for masses larger than $\SI{0.05}{\eV}$. Regardless of mass, we find that the angle between the neutrino dipole and the ecliptic plane is small, implying a near-maximal annual modulation in the bulk velocity. Along with this paper, we publicly release our simulation data, comprising more than $100$ simulations for six different neutrino masses.
}

\begin{document}
\maketitle
\flushbottom

\section{Introduction}

Precise measurements of a near-perfect black-body energy spectrum and of a power-law spectrum of temperature fluctuations in the Cosmic Microwave Background (CMB) reveal detailed information about the state of the Universe at the time of decoupling around $t=10^5\,\si{\yrs}$ \cite{cobe96,wmap11,planck18b}. There is strong but indirect evidence for another Big Bang fossil in the form of $N_\text{eff}=2.99_{-0.33}^{+0.34}$ species of fermionic particles that were relativistic when the radiation decoupled \cite{planck18a}. This is consistent with the prediction of $N_\text{eff}=3.045$ for the Cosmic Neutrino Background (CNB), consisting of three species that decoupled far earlier, at only $t=\SI{1}{\s}$ \cite{weinberg62,mangano05,desalas16}. That these particles are indeed neutrinos could be confirmed if they were found to be non-relativistic today, given the standard prediction for the present-day neutrino temperature, $T_\nu=\SI{1.68e-4}{\eV}$, and the minimum mass, $m_\nu\gtrsim\SI{0.05}{\eV}$, required by neutrino oscillations for the most massive species \cite{fukuda98,ahmad02,esteban20}. Although detecting the indirect cosmological effects of massive neutrinos is challenging, this target could soon be in reach, as suggested by improved constraints on the cosmic neutrino mass fraction \cite{palanque20,choudhury20,DES21,divalentino21}.

Direct detection of relic neutrinos will be more challenging still and is likely beyond our immediate capabilities. The Karlsruhe Tritium Neutrino Experiment (KATRIN) recently placed an upper bound of $9.7\times10^{10}$ on the local neutrino overdensity relative to the cosmic mean \cite{katrin22}, far greater than the density predicted in this paper and elsewhere. An experiment specifically designed for CNB detection has been proposed by the PTOLEMY collaboration \cite{ptolemy18,betti19,ptolemy22}. Like KATRIN, the PTOLEMY proposal aims to capture neutrinos through the inverse $\beta$-decay of tritium \cite{weinberg62,cocco07}, but with targets bound to a graphene substrate to enable a larger target mass, which has its own challenges \cite{cheipesh21,ptolemy22}. Other detection proposals rely on the net momentum transfer from the neutrino wind to macroscopic test masses \cite{opher74,stodolsky74,domcke17,shergold21}, absorption features in the cosmic ray spectrum \cite{eberle04,brdar22}, blocking of neutrino emission from de-exciting atoms due to the Pauli exclusion principle \cite{yoshimura14} or the capture of neutrinos on high-energy ion beams \cite{bauer21}. We refer to \cite{bauer22} for a detailed review of the subject.

Like the CMB, the neutrino background carries both primordial or primary perturbations and secondary gravitational perturbations imprinted by the large-scale structure at late times \cite{hu95,michney06,hannestad09,lin19,tully21}. Since neutrinos are massive particles, secondary perturbations are more significant and depend on the neutrino mass and momentum, giving the background additional structure compared to the CMB. In some cases, gravitational effects may lead to slight modifications of the expected signal and in others they open up entirely new ways of testing neutrino physics. For tritium capture experiments like PTOLEMY, the expected event rate is proportional to the local number density of neutrinos \cite{betti19}, given by the monopole moment of the phase-space distribution. If the tritium targets are polarized, PTOLEMY could measure the angular power spectrum by exploiting the dependence of the event rate on the angle between the polarization and neutrino momentum axes \cite{lisanti14}. Some proposals depend on the velocity of neutrinos in the lab frame \cite{opher74,stodolsky74,domcke17,shergold21,bauer22}, while the orientation of the dipole is important for methods that rely on periodic or angular modulation of the capture rate \cite{safdi14,lisanti14,akhmedov19}. Pauli blocking could in principle probe the momentum distribution \cite{yoshimura14,bauer22}. Additionally, gravitational perturbations may change the flavour \cite{baushev20} and helicity \cite{baym21,hernandez_molinero22,peng22,hernandez_molinero23} makeup of the neutrino background, affecting the ability of experiments like PTOLEMY to distinguish between Dirac and Majorana neutrinos.

To determine the prospects of current and future CNB detection proposals, we therefore need to model the phase-space distribution of relic neutrinos, including its higher-order directional perturbations. Previous studies have looked at the gravitational enhancement of the monopole moment due to the Milky Way \cite{ringwald04,desalas17,zhang17,mertsch19,holm23} and nearby Andromeda and Virgo \cite{mertsch19}. A very recent study also considered the gravitational influence of dark matter structures in a random $(\SI{25}{\Mpc})^3$ region on the neutrino phase-space distribution \cite{zimmer23}. Here, we expand on these works in several ways. First and foremost, we model the full six-dimensional phase-space distribution of relic neutrinos, taking into account perturbations imprinted on the neutrinos before they entered our galactic neighbourhood. Second, we use self-consistent cosmological simulations to accurately model the time evolution of the large-scale structure and the neutrino background. Third, we use an accurate non-linear treatment of massive neutrinos \cite{elbers21}, which includes the gravitational effects of the neutrinos themselves. Fourth, we model the large-scale distribution of matter within $\SIF{200\invhub}{\Mpc}$\footnote{In this expression, $h$ is defined in terms of Hubble's constant as $h \equiv H_0 / (\SI{100}{\km\per\s\per\Mpc)}$.} over the full sky, using observations from the \texttt{2M++} galaxy redshift catalogue \cite{lavaux11}. Fifth, we use a more recent estimate of the Milky Way mass from \cite{cautun19}, which is significantly lower than the value used in previous studies, depressing the effect of the Milky Way.

Using our constrained phase-space simulations, we compute the expected density, velocity, and direction of relic neutrinos, as well as expected event rates for PTOLEMY. We also study the distribution of angular anisotropies, finding that local neutrino density perturbations are anti-correlated with the projected matter distribution, due to the capture and deflection of neutrinos by massive objects along the line of sight. To facilitate future analyses of the neutrino phase-space distribution, we publicly release our simulation data alongside this paper (see Appendix~\ref{sec:opendata}). The paper is organized as follows. We describe our simulation and calibration methods in Section \ref{sec:methods}. Our main results are presented in Section \ref{sec:results}. We finally conclude in Section \ref{sec:discussion}.

\section{Methods}\label{sec:methods}

We now describe our simulation and analysis methods, starting with the details of the constrained simulations in Section \ref{sec:constrained}, our calibration procedure for applying \texttt{2M++} constraints to different neutrino cosmologies in Section \ref{sec:calibration}, and our treatment of non-linear neutrino perturbations in Section \ref{sec:reverse}.

\subsection{Constrained simulations}\label{sec:constrained}

Our analysis is based on constrained $\Lambda$CDM simulations of the local Universe. Whereas most cosmological simulations start from random initial conditions and only reproduce observations in a statistical sense, constrained simulations employ specialized initial conditions that give rise to an \emph{in silico} facsimile of the observed large-scale structure. Within the precision of the constraints, objects appear in the right relative positions and with the right dimensions, enabling a one-to-one comparison with observations. The past few years have seen constrained simulations being used for a wide range of applications and employing a variety of methods to set up the initial conditions \cite{yepes13,wang14,libeskind20,sorce21,mcalpine22,oei22}. In this paper, we use a Bayesian forward modelling approach known as `Bayesian Origin Reconstruction from Galaxies' (BORG) \cite{jasche12,jasche19,stopyra23}. This approach uses a Hamiltonian Monte Carlo algorithm to draw samples from the posterior distribution of initial conditions, given a likelihood function that connects initial conditions with observations and a Gaussian prior. The forward model consists of a Comoving Lagrangian Acceleration (COLA) code \cite{tassev13} that approximates the process of structure formation in the $\Lambda$CDM paradigm and a non-linear bias model that connects the final dark matter density field to observed galaxy positions. The Hamiltonian Monte Carlo algorithm is used to efficiently sample a high-dimensional parameter space, consisting of a grid of $256^3$ initial phases, multiple bias parameters, and the observer velocity in the CMB frame.

\begin{figure*}
    \centering
    \subfloat{
        \hspace{2em}\includegraphics{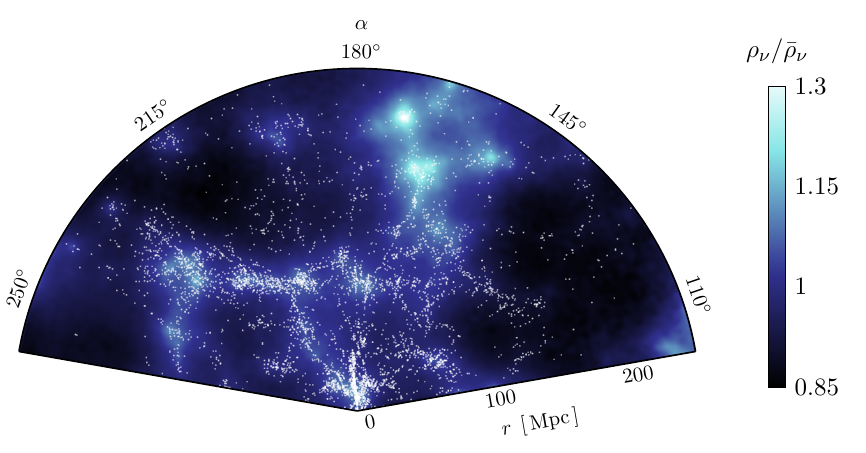}
    }~\\\vspace{-3em}
    \subfloat{
        \hspace{2em}\includegraphics{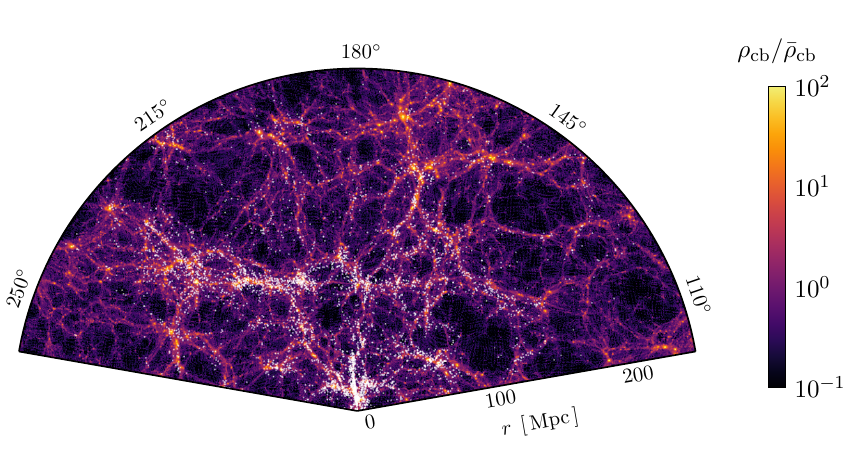}
    }
    \caption{Slices of the expected neutrino (top) and dark matter (bottom) densities with right ascension $100^\circ \leq \alpha \leq 260^\circ$ within $r\leq\SI{250}{Mpc}$, assuming $\sum m_\nu=\SI{0.06}{\eV}$. The white dots are galaxies from the \texttt{2M++} catalogue. From Earth, one would see a deficit in neutrino flux along lines of sight that intersect massive structures, due to the trapping of neutrinos in the surrounding neutrino clouds.}
	\label{fig:neutrino_and_cdm}
\end{figure*}

The constraints used in this paper are based on galaxies from the \texttt{2M++} catalogue \citep{lavaux11}. This is a catalogue of galaxy positions and redshifts, compiled from the 2MASS, 6dF, and SDSS redshift surveys, that covers the full sky out to a distance of $\SIF{200\invhub}{\Mpc}$. Previous simulations with initial conditions based on forward modelling of \texttt{2M++} galaxies include the \texttt{CSiBORG} suite \cite{bartlett21,bartlett22,desmond22,hutt22} and the \textsc{sibelius-dark} simulation \cite{mcalpine22}. We refer the reader to \cite{jasche19,stopyra23} for further details on the BORG analysis of the \texttt{2M++} catalogue. This analysis provides not only an accurate reconstruction of the three-dimensional density field in the local Universe, but also reproduces the masses of nearby clusters, with the notable exception of the Perseus-Pisces cluster for which the mass is biased low \cite{stopyra23}. This is most likely due to a systematic error in the analysis, but could perhaps also indicate an observational issue \cite{stopyra23}. Interestingly, the \textsc{sibelius-dark} simulation \cite{mcalpine22}, which is based on a similar but older BORG reconstruction, found its most massive dark matter halo at the location of Perseus. However, \textsc{sibelius-dark} was less accurate in other respects, such as the motion of the Local Group, which is important for our purposes here. Our work is based on nine draws from an earlier version of the chain described in \cite{stopyra23}, which used ten COLA steps instead of twenty, but was identical in every other respect. We therefore expect the results to be broadly consistent. After discarding an initial burn-in portion, we selected every $432$nd draw from the chain to minimize the serial correlation between consecutive draws. This sample of initial conditions allows us to estimate both the expected signal and the uncertainty in our predictions. To demonstrate the effectiveness of the constraints, we show slices of the dark matter and neutrino densities in a portion of the sky in Fig.~\ref{fig:neutrino_and_cdm}, overlaid with \texttt{2M++} galaxies (white dots). All prominent structures present in the catalogue are reproduced by the simulations, revealing the underlying dark matter filaments and surrounding neutrino clouds.

Our simulations assume periodic boundary conditions in a $(\SI{1}{\Gpc})^3$ cube, with the observer located at the centre. The \texttt{2M++} constraints mostly cover a central sphere of radius $\SI{200}{\Mpc}$ and gradually taper off beyond that. This means that sufficiently far away from the centre, the initial conditions revert to purely random fluctuations. Given that the phases are provided in the form of $256^3$ grids, the constraints only cover $\SI{4}{\Mpc}$ scales and larger. Fluctuations on smaller scales are unconstrained and purely random. Dark matter initial conditions are generated with 3LPT at $z=31$, using a modified version of \textsc{monofonIC} that adds corrections from massive neutrinos \cite{michaux20,hahn20,elbers22}, while the neutrinos themselves are generated with \textsc{fastdf}, using linear geodesic integration \cite{elbers22b}. The transfer functions are computed with \textsc{class} \cite{lesgourgues11,lesgourgues11b}.

The simulations were carried out with a version of \textsc{Gadget-4} \cite{springel20} that was modified to be bitwise reversible (see Appendix~\ref{sec:reversible}) and to add support for massive neutrinos and radiation. We use a $3^\text{rd}$-order Tree-PM algorithm for the gravity calculation. Neutrinos are followed with the $\delta f$ method to minimize shot noise, boosting the effective particle number without neglecting their non-linear evolution \cite{elbers21}. We use $N_\text{cb}=384^3$ dark matter and baryon particles\footnote{We will treat cold dark matter and baryons as a single cold fluid and refer to it as dark matter on occasion.} and $N_\nu = 384^3$ massive neutrino particles. In order to increase the sampling density of neutrinos locally, upon completion of a simulation, we isotropically inject an additional $N=224^3\sim10^7$ `spectator' neutrinos at the observer location and run the simulations backwards, allowing us to trace the neutrinos back in time through the evolving large-scale structure (see Section \ref{sec:reverse}). To ensure that the accelerations are identical in the forwards and backwards directions, spectator neutrinos contribute no forces.

A final consideration is that Milky Way-sized perturbations have a characteristic length that is much smaller than $\SI{4}{\Mpc}$. Hence, our constraints are not sufficient to guarantee the formation of a Milky Way at the centre. Since we expect the Milky Way (MW) to have a considerable effect on the neutrino background, we run two backwards versions of each simulation. Initially, neutrinos are only traced back through the large-scale structure without accounting for MW effects. In the second version, we additionally apply forces from the MW dark matter halo. Following \cite{cautun19}, we model the MW halo as an NFW profile \cite{navarro96} with a mass of $M_{200}=0.82 \times 10^{12}M_\odot$ and a concentration of $c_{200}=13.31$.\footnote{Here, $M_{200}$ is the mass contained in a spherical region of radius $R_{200}$ with a density equal to 200 times the critical density and $c_{200}=R_{200}/R_s$, with $R_s$ the scale radius of the NFW profile.} For computational simplicity, we use the uncontracted version of the model, since both versions fit the data nearly equally well. We place the centre of the NFW potential at a distance of $\SI{8}{\kpc}$ from the centre of the simulation in the direction of Sag-$\text{A}^*$. We also include the motion of the galactic centre in the CMB rest frame of the simulation, by letting the centre of the NFW potential move at a constant speed of $\SI{567}{\km/\s}$ in the direction of galactic coordinates $(l,b)=(267^\circ,29^\circ)$ \cite{tully07,tully16,planck18b}. In Section~\ref{sec:abundance}, we additionally correct for the motion of the Sun relative to the CMB, $v_\odot=\SI{369.8}{\km/\s}$ towards $(l,b)=(264^\circ,48.3^\circ)$ \cite{planck18b}, which is otherwise unresolved by the simulations. Crucially, we note that we use a more recent and considerably smaller estimate of the MW mass than that used in previous related works \cite{desalas17,mertsch19}. We therefore expect to find a smaller effect from the MW. Since we are mainly interested in the imprint of the large-scale structure, we do not include the various gaseous and stellar components of the MW, which are altogether less important than the dark matter halo itself.

\subsection{Model selection}\label{sec:calibration}

To derive constrained initial conditions with BORG, we have to assume a particular cosmological model. The constraints used in this paper were derived assuming a flat $\Lambda$CDM model with parameters $(\Omega_\text{cdm},\,\Omega_\text{b},\,h,\,A_s,\,n_s,\,\sum m_\nu) = (0.2621,\,0.04897,\,0.6766,\,2.105\times10^{-9},\,0.9665,\,0)$. Despite the fact that this model does not include massive neutrinos, we wish to run constrained simulations for different neutrino masses, without running an expensive MCMC analysis for each case. Doing this requires modifying the cosmological model slightly without altering the clustering on small scales, since otherwise the same phase information would give rise to structures that differ somewhat from the observations. We therefore take the following approach. When increasing $\sum m_\nu$, we decrease $\Omega_\text{cdm}$ such that $\Omega_\text{m} = \Omega_\text{cdm} + \Omega_\text{b} + \Omega_\nu$ is fixed. In addition, we modify the primordial scalar amplitude $A_s$, such that the non-linear power spectrum at $z=0$ is fixed at the non-linear scale $k_\text{nl}=\SI{1}{\per\Mpc}$. Note that $P_\text{cb}$, the power spectrum of cold dark matter and baryons, is the relevant power spectrum, given that halos are primarily biased with respect to the cold matter, as opposed to the total matter density \cite{ichiki11,castorina13,massara14}. To achieve this in practice, we perform a small number of calibration runs and iteratively select values of $A_s$ that satisfy this condition.

\begin{figure}
    \centering
    \subfloat{
        \includegraphics{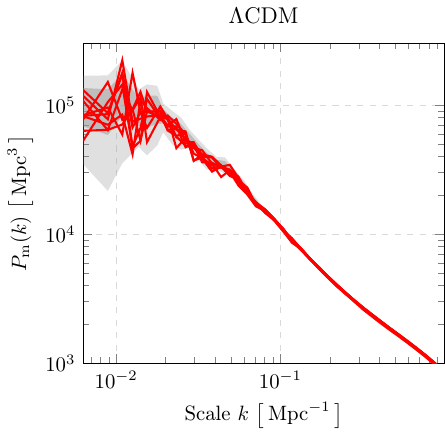}
    }
    \subfloat{
        \hspace{-1em}
        \includegraphics{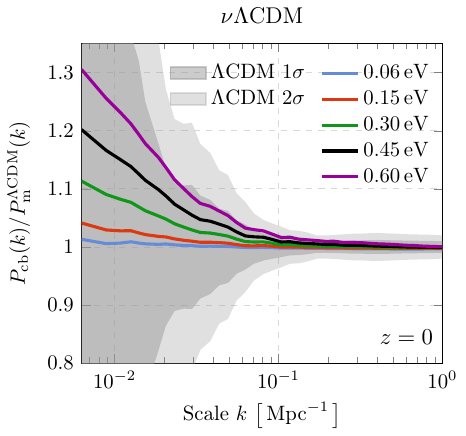}
    }
    \caption{(Left) The red lines are nine non-linear matter matter power spectra, $P_{\rm{m}}(k)$, drawn from the posterior distribution of the \texttt{2M++} reconstruction, assuming $\Lambda$CDM with massless neutrinos at $z=0$. The shaded areas represent the $1\sigma$ and $2\sigma$ deviations from the mean. The spectra are well-constrained for $\SI{0.03}{\per\Mpc}\lesssim k\lesssim \SI{1}{\per\Mpc}$, but the variance is considerable on large scales. (Right) Ratios of the non-linear cold matter power spectrum, $P_{\rm{cb}}(k)$, for different neutrino masses relative to the massless $\Lambda$CDM case, calibrated to match the constraints in the small-scale limit. The shaded areas represent the $1\sigma$ and $2\sigma$ constraints.}
	\label{fig:calibration_spectra}
\end{figure}

As noted before, the \texttt{2M++} data mostly constrain scales larger than $\SI{4}{\Mpc}$ within $\SI{200}{\Mpc}$ of the observer. As shown in Fig.~\ref{fig:calibration_spectra}, this leaves enough flexibility on large scales to accommodate neutrino masses up to $\sum m_\nu\sim\SI{0.6}{\eV}$.\footnote{We note that this breaks the agreement with CMB observations, which primarily constrain large scales. This is simply another way of stating that the combination of CMB and LSS data can rule out large neutrino masses in $\nu\Lambda$CDM, although we make no attempt to do this here.} To see this, note that the left-hand panel shows total matter power spectra, $P_\text{m}(k)$, for nine realizations assuming $\Lambda$CDM without massive neutrinos. Although the power spectrum is well-constrained on small scales, there is considerable variance on large scales ($k\lesssim\SI{0.03}{\per\Mpc}$). The right-hand panel shows the power spectrum of dark matter and baryons, $P_\text{cb}(k)$, for the calibrated models with different neutrino masses, relative to the massless case. For the largest mass considered, $\sum m_\nu=\SI{0.6}{\eV}$, the ratio is still within $1\sigma$ of the average. We also checked that the cross-correlation coefficients of the final density fields are within $1\%$ for $k\leq k_\text{nl}$ and $\sum m_\nu\leq\SI{0.3}{\eV}$ and within a few per cent for $\sum m_\nu \leq \SI{0.6}{\eV}$, indicating that the phase information is the same on large scales. Finally, we performed a visual inspection to confirm that the we recover the same large-scale structure for all neutrino masses. Hence, the outcome of this procedure is a plausible cosmological model with massive neutrinos that reproduces the \texttt{2M++} observations.

Although the resulting power spectra are compatible with the \texttt{2M++} constraints at the $1\sigma$-level, one may wonder whether the $20\%-30\%$ differences seen for $\sum m_\nu=\SI{0.6}{\eV}$ on the largest scales could still affect the results. We expect the impact of this offset to be small, because the distance travelled by neutrinos is inversely proportional to the mass, such that heavier neutrinos are less sensitive to large-scale density perturbations. Therefore, matching only the small-scale power spectrum for $\sum m_\nu=\SI{0.6}{\eV}$ is likely justified.

\begin{table}
   \centering
   \caption[caption]{Cosmological parameters for our six neutrino models, which have been calibrated such that $\Omega_\nu + \Omega_\text{cdm}$ and $P_\text{cb}(k_\text{nl})$ with $k_\text{nl}=\SI{1}{\per\Mpc}$ are fixed.}
   \begin{tabular}{ l l l l l l }
       \hline
       \hline
       $\sum m_\nu$ & Mass $m_\nu$ & $N_\nu$ & $\Omega_\nu$ & $\Omega_\text{cdm}$ & $A_s$ \\
       \hline
       $\SI{0.01}{\eV}$ & $\SI{0.01}{\eV}$ & $1$ & $2.353\times10^{-4}$ & $0.26189$ & $2.107\times10^{-9}$ \\
       $\SI{0.06}{\eV}$ & $\SI{0.06}{\eV}$ & $1$ & $1.407\times10^{-3}$ & $0.26072$ & $2.156\times10^{-9}$ \\
       $\SI{0.15}{\eV}$ & $\SI{0.05}{\eV}$ & $3$ & $3.518\times10^{-3}$ & $0.25861$ & $2.243\times10^{-9}$ \\
       $\SI{0.30}{\eV}$ & $\SI{0.10}{\eV}$ & $3$ & $7.035\times10^{-3}$ & $0.25509$ & $2.429\times10^{-9}$ \\
       $\SI{0.45}{\eV}$ & $\SI{0.15}{\eV}$ & $3$ & $1.055\times10^{-2}$ & $0.25157$ & $2.641\times10^{-9}$ \\
       $\SI{0.60}{\eV}$ & $\SI{0.20}{\eV}$ & $3$ & $1.407\times10^{-2}$ & $0.24805$ & $2.878\times10^{-9}$ \\
       \hline
       \hline
   \end{tabular}
   \label{tab:parameters}
\end{table}

Using the above procedure, we calibrate six models with different neutrino masses: four models with three degenerate neutrino species, $\sum m_\nu\in\{0.15,0.3,0.45,0.6\}\;\si{\eV}$\footnote{Hence, the individual neutrinos have masses $m_\nu\in\{0.05,0.1,0.15,0.2\}\;\si{\eV}$.}, and two models with a single neutrino species, $\sum m_\nu\in\{0.01,0.06\}\;\si{\eV}$. The relevant model parameters are given in Table~\ref{tab:parameters}. Although not strictly allowed by oscillation data, the first four models assume a degenerate neutrino mass spectrum, neglecting the mass-squared differences $\rvert\Delta m^2_{31}\rvert= \SI{2.5e-3}{\eV^2}$ and $\Delta m^2_{21}=\SI{7.4e-5}{\eV^2}$ \cite{esteban20}. Of course, the last two models are also not allowed. The penultimate case is included to examine the behaviour of very light neutrinos. The last model is included as it approximates the cosmological effects of the minimal neutrino mass case under the normal mass ordering. In each case, the intent is only to recover the correct cosmological evolution for a given neutrino mass, $m_\nu$, and for this purpose, the mass splittings have a negligible effect \cite{archidiacono20}.

\subsection{Neutrino treatment}\label{sec:reverse}

Let us now discuss our treatment of neutrino perturbations. The evolution of the phase-space distribution, $f(\mathbf{x},\mathbf{q},\tau)$, is governed by the collisionless Boltzmann equation:
\begin{align}
\frac{\partial f}{\partial\tau} + \frac{\mathrm{d}\mathbf{x}}{\mathrm{d}\tau}\cdot\nabla f + \frac{\mathrm{d}\mathbf{q}}{\mathrm{d}\tau}\cdot\nabla_{\mathbf{q}}f = 0,
\end{align}

\noindent
where $\tau$ is conformal time and $\mathbf{q}$ the neutrino momentum. We solve this equation by generating particles from a sampling distribution $g(\mathbf{x},\mathbf{q})$ and tracing their evolution through the constrained volume using the relativistic equations of motion \cite{elbers22b}
\begin{align}
	\frac{\mathrm{d}x^i}{\mathrm{d}\tau} &= \frac{q^i}{\sqrt{q^2+m^2a^2}}, \label{eq:x_eq}\\
	\frac{\mathrm{d}q_i}{\mathrm{d}\tau} &= -ma\,\nabla_i\Phi, \label{eq:q_eq}
\end{align}

\noindent
where $a$ is the scale factor and $\Phi$ the gravitational potential. The sampling distribution $g$ need not be the same as the physical distribution $f$ and can be chosen arbitrarily, subject to being normalized and the set $\{g=0 \,\land\, f\neq 0\}$ having measure zero. We model neutrino perturbations with the $\delta f$ method \cite{elbers21}, a variance reduction technique in which the phase-space distribution is decomposed as $f(\mathbf{x},\mathbf{q},\tau) = \bar{f}(q) + \delta f(\mathbf{x},\mathbf{q},\tau)$. In this approach, particle data are only used to estimate the perturbation $\delta f$ to an analytical background model $\bar{f}$, which allows sampling noise to be reduced by orders of magnitude. Let $A$ be an arbitrary phase-space statistic, such as the number or momentum density. Then, its $\delta f$ estimate is
\begin{align}
	A(\mathbf{x},\tau) &= \int\mathrm{d}^3q\left[\bar{f}(\mathbf{x},\mathbf{q},\tau) + \delta f(\mathbf{x},\mathbf{q},\tau)\right]A(\mathbf{x},\mathbf{q},\tau)\\
    &\cong \bar{A}(\tau) + \sum_{k=1}^N \frac{\delta f(\mathbf{x}_k,\mathbf{q}_k,\tau)}{g(\mathbf{x}_k,\mathbf{q}_k)}A(\mathbf{x}_k,\mathbf{q}_k,\tau) \, \delta^{(3)}(\mathbf{x}-\mathbf{x}_k), \label{eq:phase_stats}
\end{align}

\noindent
where $\bar{A}$ is the analytical background solution and we sum over particle data $\{\mathbf{x}_k,\mathbf{q}_k\}$. The Dirac function is often replaced with a spatial smoothing kernel $W(\mathbf{x}-\mathbf{x}_k)$. We can also define angular statistics. For example, the density of neutrinos at $\mathbf{x}$ with momenta oriented along the unit vector $\hat{n}$ is
\begin{align}
n_\nu(\mathbf{x},\hat{n},\tau) &= \int \mathrm{d}^3q\left[\bar{f}(\mathbf{x},\mathbf{q},\tau) + \delta f(\mathbf{x},\mathbf{q},\tau)\right]\delta^{(2)}(\mathbf{q}/\!q-\hat{n})\\
&\cong \frac{\bar{n}_\nu(\tau)}{4\pi} + \sum_{k=1}^N \frac{\delta f(\mathbf{x}_k,\mathbf{q}_k,\tau)}{g(\mathbf{x}_k,\mathbf{q}_k)}\delta^{(2)}(\mathbf{q}_k/\!q_k-\hat{n})\delta^{(3)}(\mathbf{x}-\mathbf{x}_k), \label{eq:phase_stats_flux}
\end{align}
\noindent
where $\bar{n}_\nu(\tau)$ is the mean number density and where $\delta^{(2)}(\hat{x}-\hat{y})=\delta(\cos\theta-\cos\theta')\delta(\phi-\phi')$. The fractions on the right-hand side of \eqref{eq:phase_stats} and \eqref{eq:phase_stats_flux} correspond to statistical weights, $w = \delta f/g$, which are simple to compute in practice, given that the background density $\bar{f}$ is an analytical function and that $f$ and $g$ are conserved along particle trajectories. Throughout, we use a standard Fermi-Dirac distribution, $\bar{f}(q)=(1+\exp(q/k_bT_\nu))^{-1}$, for the background model and we set $g=f$ when generating the initial conditions.

This approach is sufficient for describing the neutrino distribution on large scales, as illustrated in Fig.~\ref{fig:neutrino_and_cdm} for $\SI{0.06}{\eV}$ neutrinos. However, given the $(\SI{1}{\Gpc})^3$ ambient volume of our simulations, there is a more efficient way to estimate the properties of neutrinos incident on Earth. For this, we inject `spectator' neutrinos at the location of Earth and run our simulations backwards. For these neutrinos, we adopt an isotropic Fermi-Dirac sampling distribution $g$. We then apply our $\delta f$ logic in reverse: given the known sampling density $g$ and the background density $\bar{f}(q)$ with the momentum $q$ from the final ($z=31$) snapshot of the backwards simulation, we obtain the statistical weight $w=(\bar{f}-g)/g$. We again estimate phase-space statistics using Eq.~\eqref{eq:phase_stats}. Note that in this case, the assumed sampling distribution $g$ is not equal to the physical distribution $f$. In particular, we do not expect the distribution of local relic neutrinos to be exactly isotropic. However, the assumption of an isotropic and homogeneous Fermi-Dirac distribution at $z=31$ still allows us to use Eq.~\eqref{eq:phase_stats} to obtain physical phase-space estimates. Finally, we note that running $N$-body simulations backwards is non-trivial and we refer the reader to Appendix~\ref{sec:reversible} for details on how this is accomplished. 

\section{Results}\label{sec:results}

Having described our simulation methods, we are now in a position to discuss the results. In Section \ref{sec:abundance}, we present the expected number density, bulk velocity, and deflection angles of relic neutrinos in the Milky Way. We also compute expected event rates for PTOLEMY. In Section \ref{sec:anisotropies}, we turn to the angular distribution of neutrino anisotropies. In Section \ref{sec:cosmography}, we adopt a cosmographical perspective and look at maps of the large-scale distribution of neutrinos in the local Universe.

\begin{figure}
    \centering
    \subfloat{
        \includegraphics{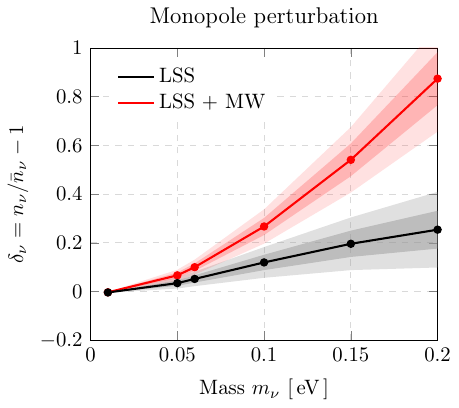}
    }
    \subfloat{
        \hspace{-1.5em}
        \includegraphics{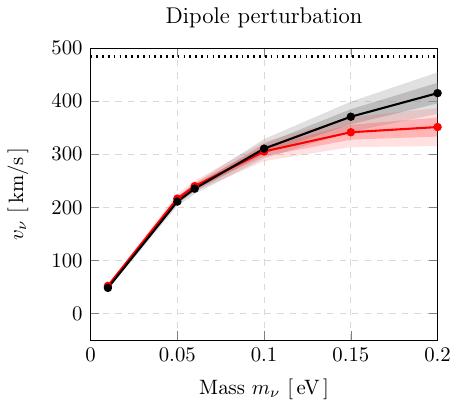}
    }

    \caption{(Left) The predicted enhancement of the local neutrino density, $\delta_\nu$, as a result of the observed large-scale structure in the local Universe (LSS, black) and the combined effect of the large-scale structure and the Milky Way (LSS $+$ MW, red). The mean and standard deviation are estimated from nine draws from the posterior distribution of the \texttt{2M++} reconstruction. (Right) The effect of the large-scale structure and Milky Way on the bulk neutrino velocity, $v_\nu$, in the CMB rest frame. The horizontal dotted line indicates the bulk velocity of CDM and baryons within $\SI{10}{\Mpc}$ of the observer.}
	\label{fig:density_bulkflow}
\end{figure}

\subsection{Local abundance and bulk motion}\label{sec:abundance}

A crucial input for relic neutrino detection efforts is the expected gravitational enhancement of the local neutrino density. Using our constrained simulations, we are able for the first time to compute the total effect of the observed large-scale structure. The result is shown in the left-hand panel of Fig.~\ref{fig:density_bulkflow}. The black line (labelled LSS) shows the effect from the large-scale structure, excluding the Milky Way, on the neutrino overdensity, $\delta_\nu=n_\nu/\bar{n}_\nu-1$, as a function of neutrino mass $m_\nu$. The error bars indicate the dispersion among the nine constrained realizations. We see that the enhancement is negligible for $m_\nu\leq \SI{0.05}{\eV}$. In fact, for the smallest mass of $\SI{0.01}{\eV}$, we find a small deficit of $\delta_\nu=-0.0038\pm0.0006$. From there, the density contrast increases approximately linearly with mass up to $\delta_\nu=0.25\pm0.08$ for $\SI{0.2}{\eV}$.

The red line shows the combined effect of the large-scale structure and the Milky Way dark matter halo (LSS + MW). The importance of the MW increases with mass, relative to the LSS. For $m_\nu=\SI{0.1}{\eV}$, they are approximately equally important. For $m_\nu=\SI{0.2}{\eV}$, the MW is responsible for three-quarters of the effect. This is a result of the decrease in free-streaming length with mass: at average speed, an unperturbed $\SI{0.01}{\eV}$ neutrino has travelled $\SI{3.1}{\Gpc}$ since $z=31$, while the number is only $\SI{200}{\Mpc}$ for $\SI{0.2}{\eV}$. As a result, lighter neutrinos are sensitive to more distant structures. We will confirm this explicitly in Section \ref{sec:anisotropies}. Taking the difference between the results with and without the MW, we find that the galactic effect is well described by $\delta_\nu^\text{MW}=27.6(m_\nu/\SI{1}{\eV})^{2.29}$. The near-quadratic scaling agrees with \cite{zhang17}, who found $\delta_\nu^\text{MW}=76.5(m_\nu/\SI{1}{\eV})^{2.21}$, but our amplitude is three times smaller. Similarly, we find significantly smaller overdensities compared to \cite{desalas17,mertsch19,holm23}. This may be partially due to the absence of gaseous and stellar Milky Way components in our simulations. However, the primary reason is most likely the more recent but lower estimate of the dark matter mass used in this work ($M_{200}=0.82 \times 10^{12}M_\odot$ here compared to $M_{200}=3.34\times 10^{12}M_\odot$ in \cite{desalas17} and $M_{200}=1.79\times10^{12}M_\odot$ in \cite{mertsch19}).\footnote{In this comparison, we converted their virial masses to masses within a spherical region containing 200 times the critical density. We also note that \cite{desalas17} used a generalized NFW profile with an additional parameter, precluding an exact one-to-one comparison.} To confirm this, we verified for one simulation that doubling the MW mass approximately restores agreement with \cite{mertsch19}. On the other hand, both amplitude and scaling are in good agreement with the recent study \cite{zimmer23}, who also point to a difference in halo properties, rather than methodology, to explain the disagreement with \cite{mertsch19}.

\begin{table}
   \centering
   \caption[caption]{Predictions for the neutrino dipole induced by the large-scale structure, compared with the measured CMB dipole from Planck \cite{planck18b}. The neutrino velocity, $v_\nu$, is the mean velocity in the CMB rest frame. The difference, $v_\odot - v_\nu$, is the Sun's motion in the neutrino frame. The angles $(l,b)$ correspond to the direction of the Sun's motion in the neutrino frame in galactic coordinates. The error is the standard deviation among nine realizations from the chain. The final row shows the velocity of CDM and baryons within $\SI{10}{\Mpc}$ of the observer.}
   \begin{tabular}{ l l l l l }
       \hline
       \hline
       Mass $m_\nu$ & \multicolumn{1}{c}{$v_\nu$} & \multicolumn{1}{c}{$v_{\odot} - v_\nu$} & \multicolumn{1}{c}{$l$} & \multicolumn{1}{c}{$b$} \\
       & \multicolumn{1}{c}{$[\,\si{\km/\s}\,]$} & \multicolumn{1}{c}{$[\,\si{\km/\s}\,]$} & \multicolumn{1}{c}{$[\,\si{\deg}\,]$} & \multicolumn{1}{c}{$[\,\si{\deg}\,]$} \\
       \hline
       CMB & $0$ & $369.8$ & $264.0$ & $\phantom{-}48.3$ \\
       \hline
       $\SI{0.01}{\eV}$ & $48.5\phantom{0}\pm1.5$ & $321.3\pm1.5$ & $263.6\pm0.5$ & $\phantom{-}48.2\pm0.1$ \\
       $\SI{0.05}{\eV}$ & $211.0\pm4.3$ & $193.8\pm5.5$ & $232.6\pm2.4$ & $\phantom{-}37.9\pm1.7$ \\
       $\SI{0.06}{\eV}$ & $235.3\pm5.0$ & $187.3\pm7.6$ & $225.7\pm2.5$ & $\phantom{-}32.7\pm1.2$ \\
       $\SI{0.10}{\eV}$ & $310.6\pm9.0$ & $193\phantom{.0}\pm15$ & $208.4\pm2.6$ & $\phantom{-}14.1\pm3.7$ \\
       $\SI{0.15}{\eV}$ & $371\phantom{.0}\pm14$ & $229\phantom{.0}\pm22$ & $199.6\pm2.5$ & $\phantom{-}1.6\phantom{0}\pm5.6$ \\
       $\SI{0.20}{\eV}$ & $415\phantom{.0}\pm20$ & $265\phantom{.0}\pm27$ & $195.0\pm2.7$ & $-4.5\phantom{0}\pm6.7$ \\
       \hline
       Matter & $484\phantom{.0}\pm83$ & $406\phantom{.0}\pm67$ & $206\phantom{.0}\pm11$ &  $-10\phantom{.0}\pm18$ \\ 
       \hline
       \hline
   \end{tabular}
   \label{tab:dipole}
\end{table}

Some detection proposals depend on the neutrino velocity in the lab frame \cite{opher74,stodolsky74,lisanti14,domcke17,shergold21,bauer22}. From our simulations, we estimate the bulk neutrino velocity $v_\nu$. Given that the simulation is carried out in the rest frame of the CMB, a value of $v_\nu = 0$ indicates that the neutrino dipole aligns with that of the CMB. We show the expected magnitude of the velocity perturbation in the right-hand panel of Fig.~\ref{fig:density_bulkflow}. As for $\delta_\nu$, the gravitational effect of the large-scale structure and Milky Way is negligible for $m_\nu=\SI{0.01}{\eV}$. The velocity perturbation increases to $\SI{211}{\km/\s}$ at $m_\nu=\SI{0.05}{\eV}$ and trends towards $\SI{415}{\km/\s}$ for $m_\nu=\SI{0.2}{\eV}$. These neutrinos are approximately at rest with respect to the bulk flow of matter in the inner $\SI{10}{\Mpc}$ of the simulation (see Table~\ref{tab:dipole}). When we include the effect of the Milky Way, the velocity appears to converge for the largest neutrino masses. Combined with the increased density perturbation, this indicates that the simulated MW and the surrounding structure are capable of trapping $\SI{0.2}{\eV}$ neutrinos in significant numbers.

In addition to the magnitude of the velocity perturbation, we can also predict its orientation. Table~\ref{tab:dipole} shows the predicted direction of the neutrino dipole, for the runs without MW, in galactic coordinates and compares it with the measured values for the CMB dipole from Planck \cite{planck18b} and the direction of the simulated matter flow within $\SI{10}{\Mpc}$ of the observer. For $\SI{0.01}{\eV}$, the predicted $1\sigma$ range of the neutrino dipole contains the measured CMB dipole. As $m_\nu$ increases to $\SI{0.2}{\eV}$, the values appear to converge towards the direction of the bulk flow of dark matter.\footnote{Note that the uncertainties are larger for the bulk dark matter velocity, because it is computed from the forward simulations, which have a much lower sampling density near the observer.} The results are broadly similar for the runs with MW. In the case of $\SI{0.01}{\eV}$, we find $(l,b)=(258.0^\circ\pm0.5^\circ,47.7^\circ\pm0.1^\circ)$, which is still very close to the CMB dipole. For $\SI{0.2}{\eV}$, the direction changes somewhat more to $(l,b)=(203.2^\circ\pm2.9^\circ,7.2^\circ\pm6.0^\circ)$. It is interesting to note that the ecliptic north pole is at $l=97^\circ,\;b=30^\circ$. This means that the neutrino dipole is close to the plane of Earth's orbit around the Sun, making an angle of $\phi \approx 10^\circ$. The Earth's orbital velocity is $v_\oplus \approx \SI{30}{\km/\s}$, producing a $(2v_\oplus / v_\nu)\cos\phi\sim 20\%$ perturbation for a typical neutrino velocity of $v_\nu=\SI{300}{\km\per\s}$. Hence, for experiments that depend on the neutrino velocity, an annual modulation may be detectable \cite{safdi14}. Finally, we note that the \textsc{sibelius-dark} simulation, which used similar techniques to set up the initial conditions, did not accurately reproduce the observed direction of the local matter flow \cite{mcalpine22}. We therefore caution that the theoretical uncertainty in the direction may be greater than the dispersion among the nine realizations given in Table~\ref{tab:dipole}.

\begin{table}
   \centering
   \caption[caption]{Predictions for the average deflection angle, $\cos\theta = (\mathbf{v}_\nu \cdot \mathbf{v}^\text{ini}_\nu) / (v_\nu v^\text{ini}_\nu)$, including the effects of the of the large-scale structure (LSS) and the Milky Way (LSS + MW). Using \eqref{eq:phase_stats}, we compute this from the backtraced particles with $\langle\cos\theta\rangle = (1+\sum_i w_i\cos\theta_i)/(1+\sum_iw_i)$, where $w_i$ is the phase-space weight of particle $i$ and $\cos\theta_i$ is its deflection angle.}
   \begin{tabular}{ l l l}
       \hline
       \hline
        & \multicolumn{1}{c}{(LSS)} & \multicolumn{1}{c}{(LSS + MW)} \\
       Mass $m_\nu$ & \multicolumn{1}{c}{$\langle\cos\theta\rangle$} & \multicolumn{1}{c}{$\langle\cos\theta\rangle$} \\
       \hline      
       $\SI{0.01}{\eV}$ & $0.999995\pm0.000002$ & $0.999987\pm0.000003$ \\
       $\SI{0.05}{\eV}$ & $0.99806\phantom{0}\pm0.00072$ & $0.99482\phantom{0}\pm0.00084$ \\
       $\SI{0.06}{\eV}$ & $0.9965\phantom{00}\pm0.0013$ & $0.9905\phantom{00}\pm0.0015$ \\
       $\SI{0.10}{\eV}$ & $0.9847\phantom{00}\pm0.0058$ & $0.9542\phantom{00}\pm0.0058$ \\
       $\SI{0.15}{\eV}$ & $0.958\phantom{000}\pm0.016$ & $0.869\phantom{000}\pm0.013$ \\
       $\SI{0.20}{\eV}$ & $0.923\phantom{000}\pm0.029$ & $0.754\phantom{000}\pm0.018$ \\
       \hline
       \hline
   \end{tabular}
   \label{tab:deflection_angles}
\end{table}

A related quantity to the velocity perturbation is the deflection angle between the initial and final velocities, $\cos\theta = (\mathbf{v}_\nu \cdot \mathbf{v}^\text{ini}_\nu) / (v_\nu v^\text{ini}_\nu)$. For non-relativistic neutrinos, the gravitational effect on the spin is negligible, such that a deflection of the momentum vector by an angle $\theta$ implies a change in the helicity from $\pm1$ to $\pm\cos\theta$, with a probability $P=1/2-\cos\theta/2$ of observing a reversed spin \cite{baym21}. It has recently been argued that the gravitational effect of the Virgo Supercluster might result in large deflection angles, significantly altering the helicity makeup of the neutrino background \cite{hernandez_molinero23}. These authors compute deflection angles for neutrinos in halos of a similar mass to Virgo, $M=1.48\times 10^{15}M_\odot$, finding an average of $\langle\cos\theta\rangle=0.54 - 0.60$ for $m_\nu=\SI{0.05}{\eV}$. Using our constrained simulations, which include Virgo, we can estimate directly the effect that the large-scale structure has on neutrinos that arrive on Earth. We give the average for different neutrino masses and for the cases with and without Milky Way in Table.~\ref{tab:deflection_angles}. For $\SI{0.05}{\eV}$, we find $\langle\cos\theta\rangle=0.99482\pm0.00084$, when including the Milky Way. Given that the deflection is even smaller for lighter neutrinos, we expect the effect of gravitational deflection to be negligible for the minimal neutrino mass case, $\sum m_\nu=\SI{0.06}{\eV}$.

Gravitational clustering also has the potential to alter the flavour composition of the local neutrino background \cite{baushev20}. The mass eigenstates $\nu_i$ considered so far are superpositions of flavour eigenstates $\nu_\alpha$, with $\alpha=e,\mu,\tau$, for electron, muon, and tau neutrinos. The two bases are related by the unitary Pontecorvo-Maki-Nakagawa-Sakata (PMNS) matrix $U_{\alpha i}$ \cite{pontecorvo68,maki62}. The flavour composition could be altered, since the degree of clustering depends on mass. For instance, assuming the mass ordering is normal, the contribution of $\nu_e$ to the heaviest mass state $\nu_3$ is only $\rvert U_{e3}\rvert^2 = 2.3\%$. Therefore, if $\nu_3$ is much more strongly clustered than $\nu_1$ and $\nu_2$, most relic neutrinos on Earth would be $\nu_\mu$ or $\nu_\tau$. For this effect to be large, the masses must be hierarchical ($m_1\ll m_3$ or $m_3\ll m_1$), which requires $m_\nu\lesssim\SI{0.1}{\eV}$. Fig.~\ref{fig:density_bulkflow} shows that the differences in the density contrast $\delta_\nu$ are then small, which implies that the fraction of $\nu_e$ is not significantly altered from its primordial value of $1/3$. We nevertheless incorporate this effect in the calculation below.

\begin{table}
   \centering
   \caption[caption]{Predicted number of events per year for PTOLEMY, including the effects from the large-scale structure (LSS) and the Milky Way (LSS + MW), for Dirac and Majorana neutrinos. We give the results for the individual mass states, with \eqref{eq:summed_event_rate} giving the total rate. The uncertainty corresponds to the $1\sigma$ dispersion among nine realizations from the chain.}
   \begin{tabular}{ l l l l l }
       \hline
       \hline
        & \multicolumn{2}{c}{(LSS)} & \multicolumn{2}{c}{(LSS + MW)} \\
       Mass $m_{\nu}$ & \multicolumn{1}{c}{$\Gamma_{i,\text{CNB}}^D$ $\left[\,\si{\per\yr}\,\right]$} & \multicolumn{1}{c}{$\Gamma_{i,\text{CNB}}^M$ $\left[\,\si{\per\yr}\,\right]$} & \multicolumn{1}{c}{$\Gamma_{i,\text{CNB}}^D$ $\left[\,\si{\per\yr}\,\right]$} & \multicolumn{1}{c}{$\Gamma_{i,\text{CNB}}^M$ $\left[\,\si{\per\yr}\,\right]$} \\
       \hline
       $\SI{0.01}{\eV}$ & $4.042\pm0.002$ & $8.075\pm0.005$ & $4.045\pm0.002$ & $8.080\pm0.005$ \\
       $\SI{0.05}{\eV}$ & $4.20\phantom{0}\pm0.05$ & $8.39\phantom{0}\pm0.09$ & $4.33\phantom{0}\pm0.05$ & $8.65\phantom{0}\pm0.09$ \\
       $\SI{0.06}{\eV}$ & $4.27\phantom{0}\pm0.06$ & $8.53\phantom{0}\pm0.12$ & $4.46\phantom{0}\pm0.06$ & $8.92\phantom{0}\pm0.13$ \\
       $\SI{0.10}{\eV}$ & $4.54\phantom{0}\pm0.13$ & $9.08\phantom{0}\pm0.26$ & $5.14\phantom{0}\pm0.14$ & $10.27\pm0.29$ \\
       $\SI{0.15}{\eV}$ & $4.85\phantom{0}\pm0.22$ & $9.70\phantom{0}\pm0.44$ & $6.25\phantom{0}\pm0.27$ & $12.49\pm0.54$ \\
       $\SI{0.20}{\eV}$ & $5.09\phantom{0}\pm0.32$ & $10.17\pm0.63$ & $7.60\phantom{0}\pm0.44$ & $15.19\pm0.88$ \\
       \hline
       \hline
   \end{tabular}
   \label{tab:ptolemy}
\end{table}

We now have the necessary ingredients to compute the expected event rate for an experiment like PTOLEMY. The CNB capture rate,
\begin{align}
\Gamma_\text{CNB} &= \sum_{i=1}^{N_\nu} \Gamma_{i,\text{CNB}}\rvert U_{ei}\rvert^2, \label{eq:summed_event_rate}
\end{align}
\noindent
is summed over all mass eigenstates that exceed the energy threshold of the experiment, weighted by the PMNS mixing elements, $U_{ei}$. The event rate for mass state $\nu_i$ is given by \cite{long14}
\begin{align}
\Gamma_{i,\text{CNB}} &= N \bar{\sigma} \left[n^+_iA_i^+   + n_i^- A_i^-\right], \label{eq:ptolemy}
\end{align}

\noindent
where $N$ is the number of targets, $\bar{\sigma}$ is the average cross section, $n_i^\pm$ are the number densities for the two spin states, $A^{\pm}_i = 1\mp v_i/c$ is a spin-dependent factor, and $v_i$ is the velocity of the mass eigenstate. As discussed, gravitational deflection by an angle $\theta$ reverses the spin with probability $P=1/2-\cos\theta/2$. The number densities for both spin states are then given by
\begin{align}
n^{\pm}_i &= n_i\left[\frac{1}{2} \pm \frac{1}{2}\langle\cos\theta\rangle_i\right].
\end{align}

\noindent
In the absence of clustering and deflection, $\langle\cos\theta\rangle_i=1$, such that $n^+_{i}=n_i=\bar{n}$ and $n^-_{i}=0$ for Dirac neutrinos. For Majorana neutrinos, the densities are both equal to the mean: $n^+_{i}=n^-_{i}=\bar{n}$. Consequently, for non-relativistic neutrinos with $A^\pm_i=1$, the expected signal is twice as large in the Majorana case. If we allow for gravitational effects, we instead obtain
\begin{align}
\Gamma^D_\text{CNB} &= N\bar{\sigma}\sum_{i=1}^{N_\nu} \rvert U_{ei}\rvert^2 \left[1 + \langle\cos\theta\rangle_i \frac{v_i}{c}\right] n_i,\\
\Gamma^M_\text{CNB} &= N\bar{\sigma}\sum_{i=1}^{N_\nu} \rvert U_{ei}\rvert^2 2n_i,
\end{align}

\noindent
for the Dirac and Majorana cases, respectively. Plugging in the number $N=\SI{100}{\g} / m_{{}^3\text{H}}$ of tritium atoms for PTOLEMY \cite{ptolemy18} and the average cross section $\bar{\sigma}=\SI{3.834e-45}{\cm^2}$ from \cite{long14}, and a mean number density of $\bar{n}=\SI{56}{\cm^{-3}}$ per degree of freedom, we obtain the event rates in Table~\ref{tab:ptolemy}. We report the values for the individual mass eigenstates. Comparing the most and least massive cases, we see that gravitational clustering only has a marginal effect, boosting the capture rate by less than a factor of two. For each mass, we predict a factor $\sim2$ difference between the Dirac and Majorana cases. Let us now compute the total event rate for the minimal neutrino mass case, using $\rvert U_{ei}\rvert^2=(0.678,0.299,0.023)$ \cite{esteban20}. We assume that only the heaviest neutrinos with $m_\nu=\SI{0.05}{\eV}$ ($\nu_3$ under the normal ordering or $\nu_1$ and $\nu_2$ under the inverted ordering) would produce peaks in the electron energy spectrum far enough beyond the $\beta$-decay endpoint to be detected by PTOLEMY with a reasonable energy resolution \cite{betti19}. For the normal ordering, we then find $\Gamma_\text{CNB} \approx \SI{0.1}{\per\yr}$ (Dirac) or $\SI{0.2}{\per\yr}$ (Majorana), while $\Gamma_\text{CNB}\approx \SI{4}{\per\yr}$ (Dirac) or $\SI{8}{\per\yr}$ (Majorana) for the inverted ordering.

\subsection{Angular anisotropies}\label{sec:anisotropies}

\begin{figure}
    \centering
    \subfloat{
	   \includegraphics[scale=.97]{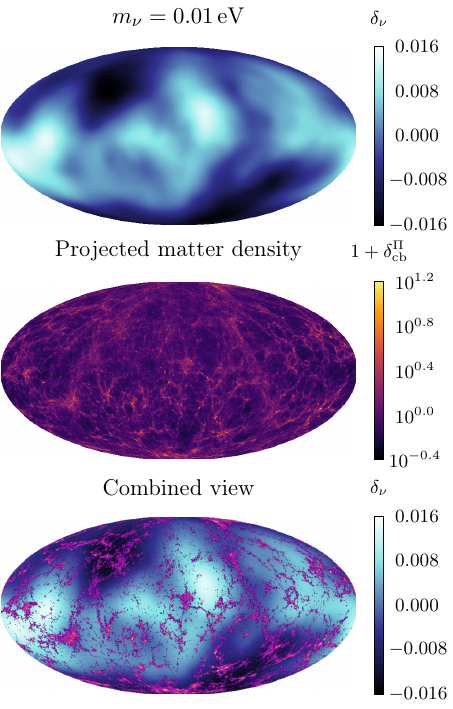}
	}
    \subfloat{
	   \includegraphics[scale=.97]{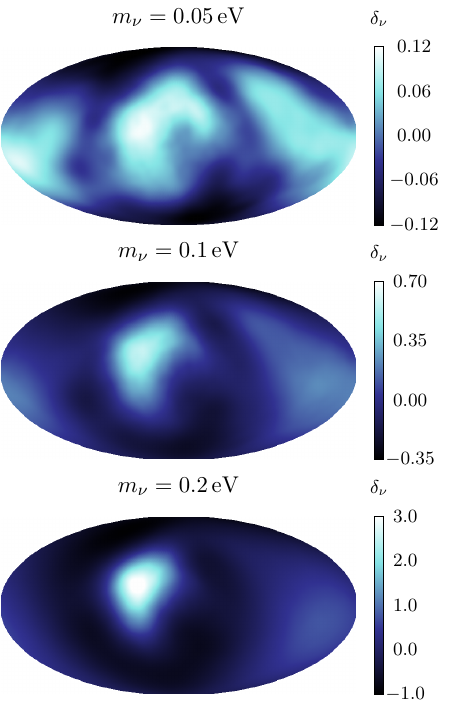}
	}
    \caption{Angular anisotropies in the neutrino number density contrast, $\delta_\nu$, for $m_\nu=\SI{0.01}{\eV}$ (top left) and for $m_\nu\in\{0.05,0.1,0.2\}\,\si{\eV}$ (right). In all cases, we subtract the monopole and dipole moments and smooth over $3^\circ$ scales. We also show the projected dark matter and baryon density, $\delta^\Pi_\text{cb}$, within $\SI{200}{\Mpc}$ of the observer, both separately (middle left) and overlaid on the top of the neutrino density for $m_\nu=\SI{0.01}{\eV}$ (bottom left). We observe that the projected dark matter density and the local neutrino density are anti-correlated on the sky. Except for the projected matter density, the maps are all based on backtraced particles.}
	\label{fig:skymaps}
\end{figure}

Having presented our results for the monopole and dipole moments, we now turn to higher-order moments of the neutrino distribution. Fig.~\ref{fig:skymaps} presents maps of the predicted angular anisotropies in the number density, $\delta_\nu(\theta,\phi) = n_\nu(\theta,\phi)/(\bar{n}_\nu/4\pi)$, for four different masses, after subtracting the monopole and dipole perturbations. The maps show relative variations in the neutrino density for individual mass eigenstates, computed via equation \eqref{eq:phase_stats_flux} by adding the weights of backtraced particles along each direction. As discussed in Section~\ref{sec:reverse}, these particles represent an ergodic ensemble of neutrino paths with weights that correct for the isotropic sampling distribution. In \cite{elbers22b}, it is shown that the statistical properties of such weighted particle ensembles are consistent with the transfer functions obtained from an Eulerian fluid calculation at the linear level. One advantage of the particle-based treatment, however, is its ability to describe the non-linear growth of neutrino perturbations, which becomes important when the neutrino mass is large, as discussed below.

Each map is averaged over nine realizations from the \texttt{2M++} reconstruction. The top-left panel of Fig.~\ref{fig:skymaps} shows the map for $m_\nu=\SI{0.01}{\eV}$ and the right-hand panels show maps for $m_\nu\in\{0.05,0.1,0.2\}\;\si{\eV}$. First of all, we observe that the magnitude of the perturbations strongly depends on mass: they are $\mathcal{O}(10^{-2})$ for $m_\nu=\SI{0.01}{\eV}$ and $\mathcal{O}(1)$ for $m_\nu=\SI{0.2}{\eV}$. We also see that the largest neutrino mass maps have large-scale perturbations that are suppressed, relative to small-scale perturbations, for the smaller neutrino masses. The middle-left panel shows the projected density of dark matter and baryons,
\begin{align}
1 + \delta^\Pi_\text{cb}(\theta,\phi,R_\text{max}) = \frac{\int_{0}^{R_\text{max}}\rho_\text{cb}(\mathbf{r})\,\mathrm{d}r}{\int_{0}^{R_\text{max}}\bar{\rho}_\text{cb}(\mathbf{r})\,\mathrm{d}r}, \label{eq:def_proj}
\end{align}

\noindent
up to a distance of $R_\text{max}=\SI{200}{\Mpc}$ from the observer. Comparing this with the neutrino maps, we find that distant matter fluctuations are anti-correlated with local neutrino fluctuations. This can be seen more clearly in the bottom-left panel, in which the projected matter perturbations are overlaid on the neutrino perturbations for $m_\nu=\SI{0.01}{\eV}$. The anti-correlation is much more evident for smaller neutrino masses.

\begin{figure}
    \centering
    \includegraphics{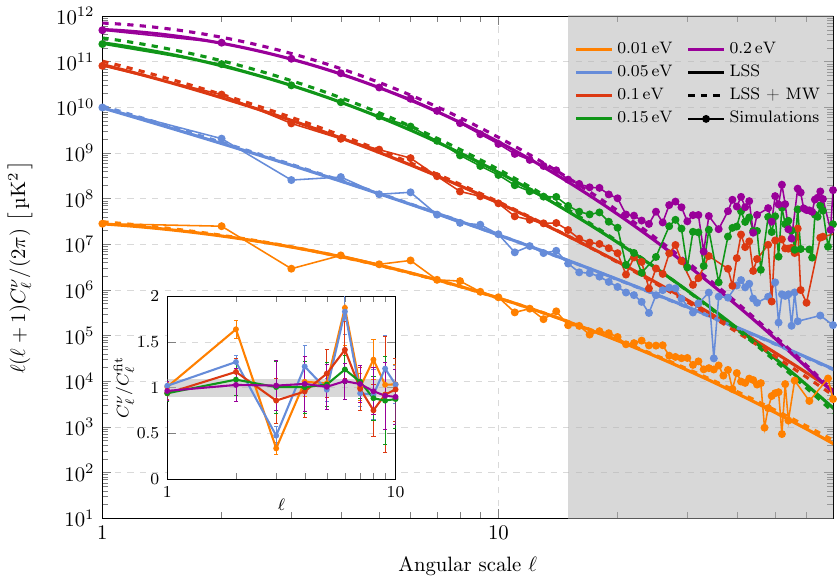}
    \caption{Angular neutrino temperature power spectra, $C^\nu_\ell$, for different masses. We fit a smooth spectrum, $C_\ell^\text{fit} = \exp(c_1 + c_2\log\ell + c_3\log^2\ell)$, up to $\ell_\text{max}=15$ for the simulations with and without a Milky Way (dashed and solid lines, respectively). To avoid clutter, we only show the data for the simulations without MW. The inset graph zooms in on the first ten multipoles, showing the data relative to the fit. The grey error bar represents $\pm10\%$. The oscillatory perturbations arise from the imprint of dark matter perturbations on the neutrino background and can ultimately be traced to cosmic variance in the matter distribution.}
	\label{fig:spectra}
\end{figure}

\begin{figure}
    \centering
    \includegraphics{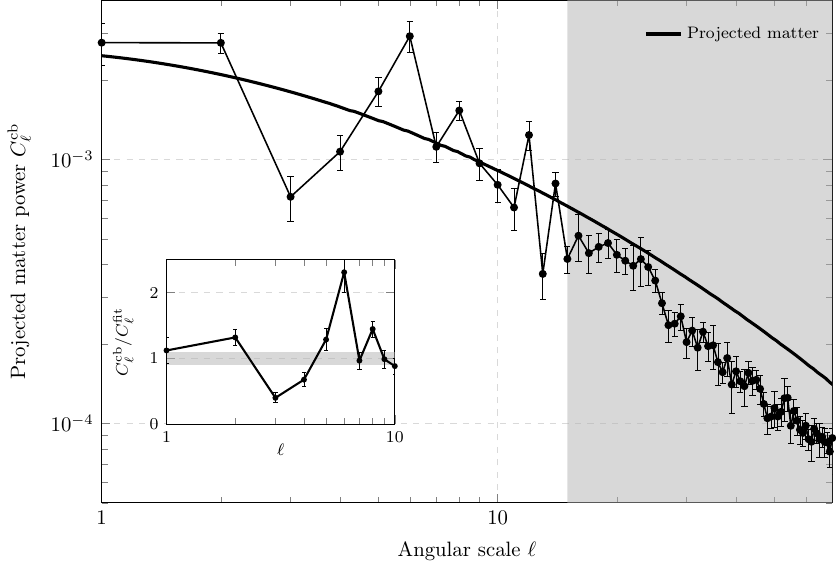}
    \caption{Dimensionless angular power spectrum of the projected CDM and baryon density contrast, $C^\text{cb}_\ell$, out to $\SI{200}{\Mpc}$ for $\SI{0.01}{\eV}$. By construction, the results are similar for different neutrino masses. We fit a smooth spectrum, $C_\ell^\text{fit} = \exp(c_1 + c_2\log\ell + c_3\log^2\ell)$, up to $\ell_\text{max}=15$ to the simulation spectrum (thick solid curve). The inset graph zooms in on the lowest-order multipoles, showing the data relative to the fit. The grey error bar represents $\pm10\%$.}
	\label{fig:spectra_2}
\end{figure}

Next, we compute angular power spectra, $C^\nu_\ell$, from the neutrino overdensity maps. To compare our results with other works \cite{hannestad09,tully21,zimmer23}, we convert the spectra to temperature units by assuming that $\delta T_\nu/\bar{T}_\nu \sim \delta n_\nu/3\bar{n}_\nu$.\footnote{This follows from the idealized result, $n_\nu = 3\zeta(3)T_\nu^3/2\pi^2$, for the Fermi-Dirac distribution, even though the actual momentum distribution of clustered neutrinos is heavily perturbed.} In Fig.~\ref{fig:spectra}, we show the results for five different masses, averaging over nine realizations from the chain. To uncover the perturbations imprinted by the large-scale structure, we fit smooth spectra of the form
\begin{align}
C^\text{fit}_\ell = \exp\left[c_1 + c_2\log\ell + c_3(\log\ell)^2\right], \label{eq:smooth_fit}
\end{align}

\noindent
to the simulation predictions, restricting to the multipoles with $1\leq\ell\leq15$, since higher-order multipoles are noisy and poorly constrained. The thick curves in Fig.~\ref{fig:spectra} correspond to these fits, with the solid and dashed lines indicating the LSS-only and combined LSS + MW results, respectively. As expected from the previous section, the effect of the MW is most pronounced for the largest neutrino masses and the lowest-order multipoles. The difference between the dashed and solid curves is negligible for $m_\nu\leq\SI{0.05}{\eV}$, but clearly visible for $m_\nu=\SI{0.2}{\eV}$. We compute our maps in the rest frame of the simulations, without accounting for observer motion. Therefore, Fig.~\ref{fig:spectra} shows the intrinsic dipole moment $(\ell=1)$ arising from large-scale matter fluctuations. The value is orders of magnitude larger than the intrinsic dipole expected for massless tracers like the CMB \cite{roldan16,meerburg17,silveira20}. This is consistent with the behaviour seen in Fig.~\ref{fig:spectra}, which shows that low-multipole perturbations are strongly enhanced for heavier neutrinos.

Our results differ substantially from \cite{zimmer23}, who compute a range of temperature power spectra for $m_\nu=\SI{0.1}{\eV}$ using different $(\SI{25}{\Mpc})^3$ simulations. We find a slope that is much steeper and an amplitude at low multipoles that is greater. This could be due to the absence of large-scale structure in their simulations, explaining the lack of power at low multipoles. Our results are in good agreement with the linear theory calculations of \cite{tully21} for $m_\nu < \SI{0.1}{\eV}$. For $\SI{0.1}{\eV}$, the normalization at low multipoles agrees, but we predict significantly more power beyond $\ell\geq10$, where the linear calculation likely breaks down. Similarly, although our definition of the neutrino temperature power spectrum is somewhat different from \cite{hannestad09}, given that we do not define a power spectrum for each momentum bin separately but show the pointwise integrated result, we obtain at least qualitative agreement with their linear calculations for masses $m_\nu\leq\SI{0.1}{\eV}$, the largest mass considered amenable to their analysis. These authors model the gravitational deflection of neutrinos with a lensing potential, similar to what is done for the CMB \cite{lewis06}. A key difference between our results and the linear theory calculations \cite{hannestad09,tully21} is the presence of oscillatory perturbations around the smooth spectra in Fig.~\ref{fig:spectra}, which are much larger than the predicted lensing effect in \cite{hannestad09}. This can be seen more clearly in the inset graph, which zooms in on the lowest-order multipoles ($\ell\leq10$) and shows the simulation predictions relative to the smooth fits. The perturbations depend sensitively on mass, being most prominent for $\SI{0.01}{\eV}$ and nearly absent for $\SI{0.2}{\eV}$.

\begin{figure}
    \centering
    \subfloat{
        \includegraphics{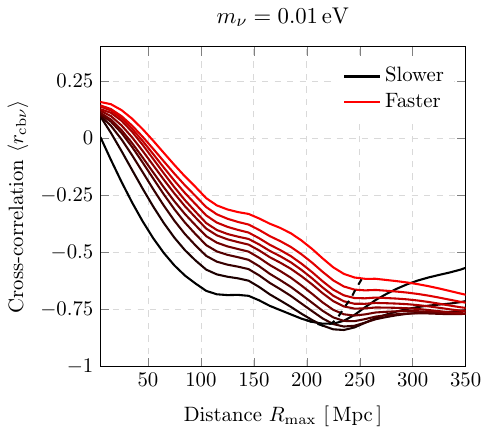}
    }
    \subfloat{
        \hspace{-1em}
        \includegraphics{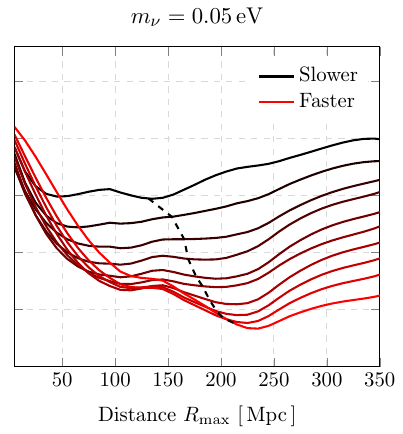}
    }
    \caption{Cross-correlation coefficient, $r_{\text{cb}\nu}(\ell) = C^{\text{cb}\nu}_\ell/(C^\nu_\ell C^\text{cb}_\ell)^{1/2}$, between the local neutrino density and the projected CDM and baryon density, as a function of the maximum projected distance $R_\text{max}$, for $m_\nu\in\{0.01,0.05\}\;\si{\eV}$, split into ten equal-sized neutrino momentum bins. The coefficients are averaged over the multipoles $1\leq\ell\leq10$ and the curves are smoothed with a Savitzky-Golay filter. The dashed line indicates the locus of the barycentre of each curve, indicating that the sensitivity shifts to larger distances for faster neutrinos. Note that the overall momentum range is much wider for the $\SI{0.01}{\eV}$ case.}
	\label{fig:binned_spectra}
\end{figure}

The origin of these perturbations becomes clear when we plot the angular power spectrum, $C^\text{cb}_\ell$, of the projected CDM and baryon density up to $\SI{200}{\Mpc}$, in Fig~\ref{fig:spectra_2}. In this case, we compute a dimensionless power spectrum directly from the maps of the projected density contrast, $\delta^\Pi_\text{cb}(\theta,\phi)$, defined in Eq.~\eqref{eq:def_proj}. Fitting a smooth power spectrum \eqref{eq:smooth_fit} in the same way as for the neutrinos, reveals the same oscillatory perturbations. This suggests that cosmic variance in the matter density field is imprinted on the local neutrino background if the neutrino mass is sufficiently small. To confirm this explicitly, we compute the cross-correlation coefficient, $r_{\text{cb}\nu}(\ell) = C^{\text{cb}\nu}_\ell/(C^\nu_\ell C^\text{cb}_\ell)^{1/2}$, between the local neutrino density and the projected dark matter and baryon density, as a function of the maximum projected distance $R_\text{max}$. The results, averaged over the lowest-order multipoles, $1\leq\ell\leq10$, and smoothed with a Savitzky-Golay filter, are shown in Fig.~\ref{fig:binned_spectra}. We additionally split the results into ten equal-sized neutrino momentum bins, with redder curves indicating faster neutrinos. For both neutrino masses shown, $m_\nu=\SI{0.01}{\eV}$ (left) and $\SI{0.05}{\eV}$ (right), there is a strong anti-correlation that peaks around $r_{\text{cb}\nu}=-0.8$. In both cases, faster neutrinos are sensitive to more distant matter fluctuations. To emphasize this point, we indicate the locus of the barycentre of each curve by a black dashed line.

Note that $r_{\text{cb}\nu}$ trends upwards as $R_\text{max}$ decreases, eventually becoming positive for the fastest neutrinos. This might be explained by the gravitational attraction of neutrinos to positive density perturbations close to the observer. In this case, a positive correlation should be expected. In line with expectation, the distance at which the correlation becomes positive increases with neutrino momentum. Interestingly, the anti-correlation becomes weaker with neutrino momentum for $\SI{0.01}{\eV}$ and stronger with neutrino momentum for $\SI{0.05}{\eV}$. A simple explanation for this could be that the anti-correlation begins trending upwards earlier for faster neutrinos, causing a reversal in the trend, as can be seen for $R_\text{max}<\SI{100}{\Mpc}$ in the case of $m_\nu=\SI{0.05}{\eV}$. For $m_\nu=\SI{0.01}{\eV}$, this reversal may only happen at distances that are not constrained by the \texttt{2M++} data underlying our simulations.

Just before this paper was submitted, a related study appeared in which neutrino anisotropy maps are analysed for different random configurations of dark matter halos in a $(\SI{25}{\Mpc})^3$ volume \cite{zimmer23}. For some configurations, they report positive or negative correlations between the neutrino and projected dark matter densities. Overall, the ensemble average of cross-power spectra is consistent with zero. Taking into account the smaller volume of the simulations, this can probably be understood in terms of the aforementioned transition from positive to negative correlations close to the observer.

\subsection{Cosmography}\label{sec:cosmography}

\begin{figure*}
	\normalsize
	\centering
	\subfloat{
		\includegraphics[scale=1]{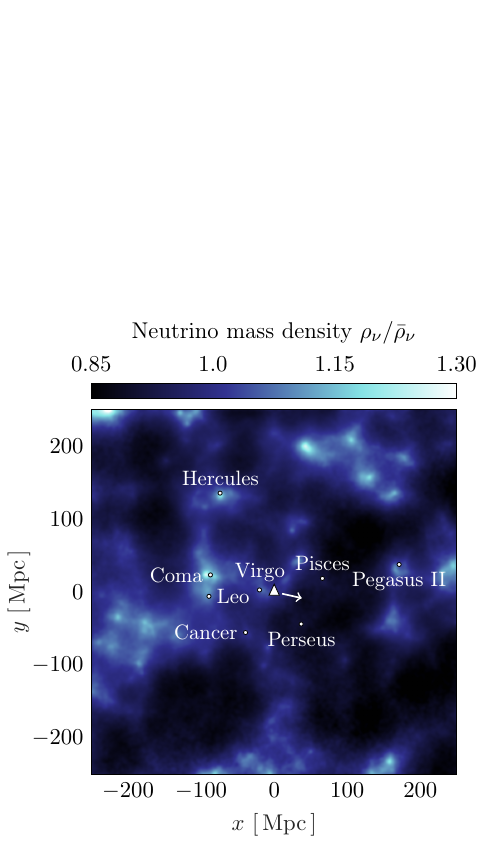}
	}
    \subfloat{
        \hspace{-1em}
		\includegraphics[scale=1]{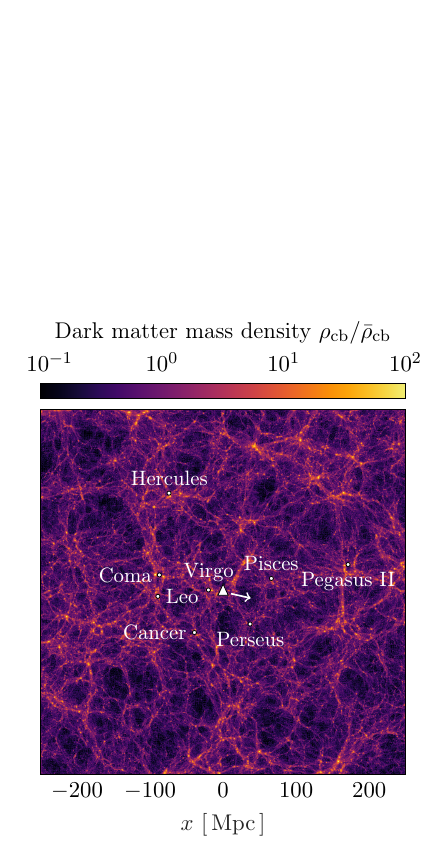}
	}\vspace{-0.5em}
	\caption{Slice of the projected neutrino (left) and dark matter (right) mass density, with thickness of $\SI{60}{Mpc}$, containing the Local Group and nearby clusters, for a species with mass $m_\nu=\SI{0.06}{\eV}$. The location of the Milky Way is indicated by a white triangle. The arrow indicates the direction of the relative neutrino velocity. In terms of the comoving distance $r$, right ascension $\alpha$, and declination $\delta$, the coordinates are $(x,y)=r\cos\delta(\cos\alpha,\sin\alpha)$.}
	\label{fig:density_plots}
\end{figure*}

In this section, we make a first attempt at neutrino cosmography. Given the limited resolution of our simulations, we focus on one illustrative example and run a higher-resolution constrained simulation with $N_\nu=N_\text{cb}=1024^3$ particles for $\sum m_\nu=\SI{0.06}{\eV}$. In Fig.~\ref{fig:density_plots}, we present maps of the neutrino density (left) and dark matter and baryon density (right), in a slice of $500\times500\times\SI{60}{\Mpc}$ that includes the Local Group and several well-known clusters. A few striking observations can be made. First of all, the large-scale neutrino and dark matter densities are positively correlated. This explains the anti-correlation seen in the previous section. Relic neutrinos that are captured by massive objects form localized clouds. Hence, while they are visible from the hypothetical viewpoint\footnote{The viewpoint of a distant observer looking at the Milky Way in its cosmic environment. One might call this the Archimedean viewpoint, based on Archimedes' claim that he could lift the Earth given only a fulcrum and a place to stand.} depicted in Fig.~\ref{fig:density_plots}, they would not be seen from Earth along lines of sight that intersect those structures.

After plotting the locations of several famous galaxy clusters, we find massive dark matter structures associated with each of them. Surrounding most of these structures, we also identify neutrino clouds that stretch over $\SI{10}{\Mpc}$ scales and reach central overdensities of $30\%$. Two interesting exceptions are the Perseus and Pisces clusters, which lie close to the Taurus void \cite{meusinger20} and appear to inhabit a large region that is deficient in neutrinos (a `glade' in the neutrino cloudscape). Although we see some collapsed dark matter structures at their locations, these are more dispersed compared to other clusters. This could be due to a failure of the constrained simulations to model the Perseus-Pisces wall accurately \cite{stopyra23}.

The Milky Way is marked by a white triangle, located along a filament that stretches towards the Virgo cluster. For this neutrino mass, $m_\nu=\SI{0.06}{\eV}$, we appear to inhabit a region with a large-scale neutrino overdensity that is not due to the Milky Way. It was this large-scale modulation of the neutrino density that originally motivated our investigation. Its effect was shown in Fig.~\ref{fig:density_bulkflow} as a function of mass. For $\SI{0.01}{\eV}$, we predicted a small neutrino deficit. We now see that this could be due to our proximity to the Taurus/Perseus-Pisces glade. Hence, the local neutrino density depends on the interplay between the overdensities associated with Virgo and the Local Group and nearby underdensities. The direction of the neutrino dipole is indicated by a white arrow. It points away from the overdense region around the Coma cluster, which is consistent with our motion towards the Shapley Supercluster and the Great Attractor \cite{lyndenbell88}. Correspondingly, it points towards an underdense region known as the Dipole Repeller, which causes an apparent repulsion \cite{hoffman17}. In short, the behaviour of the CNB dipole is similar to that of the CMB when the neutrino mass is small, consistent with our findings in Section \ref{sec:abundance}. 

\section{Conclusion}\label{sec:discussion}

Direct detection of the Cosmic Neutrino Background (CNB) remains one of the great challenges in cosmology. In this paper, we have analysed the gravitational effects of the large-scale structure and the Milky Way on the local neutrino background. Through the use of the `BORG' framework for Bayesian forward modelling of large-scale structure observations \cite{jasche12,jasche19}, we have carried out constrained simulations of the local Universe for different neutrino cosmologies with masses between $\sum m_\nu=\SI{0.01}{\eV}$ and $\sum m_\nu=\SI{0.6}{\eV}$. The constraints are based on the \texttt{2M++} catalogue \cite{lavaux11}, which maps the local Universe out to a distance of $\SIF{200\invhub}{\Mpc}$. We account for the Milky Way dark matter halo, using an updated estimate of the mass from \cite{cautun19}. By tracing neutrinos back through the galaxy and large-scale structure with a bitwise reversible version of the $N$-body code \textsc{Gadget-4} \cite{springel20}, keeping track of phase-space density perturbations, we compute statistics of the expected neutrino flux. Our results suggest that the gravitational clustering of neutrinos due to the large-scale structure is not negligible compared to the effect of the Milky Way, with both contributing about half of the total effect for $\SI{0.1}{\eV}$ neutrinos.

Despite the inclusion of the large-scale structure, we find smaller overdensities compared to earlier studies \cite{desalas17,zhang17,mertsch19,holm23}. We attribute this to a decrease in recent estimates of the Milky Way halo mass. We therefore predict only marginal increases in the event rates for tritium capture experiments like PTOLEMY \cite{ptolemy18,betti19,ptolemy22}. Additionally, we also predict a smaller impact of gravitational deflection on the helicity distribution of the neutrino background compared to \cite{hernandez_molinero23}, due to our distance from the centre of the Virgo cluster. As a result, the difference between the event rates for Dirac and Majorana neutrinos is slightly smaller, though still close to $100\%$ in most cases. Similarly, we also predict a smaller impact on the flavour composition compared to \cite{baushev20}, with an electron-neutrino fraction that is close to $1/3$ even for hierarchical masses. We also make predictions for the neutrino dipole. In the limit of very small neutrino masses, $m_\nu\leq\SI{0.01}{\eV}$, we recover the CMB result with a dipole that corresponds to Solar motion towards $(l,b)=(264^\circ,\,48^\circ)$ at a relative velocity of around $\SI{300}{\km/\s}$. The velocities are significantly perturbed for larger masses and the dipole direction shifts, but remains nearly parallel to the ecliptic plane. This implies a near-maximal annual modulation in the neutrino velocity throughout Earth's orbit around the Sun.

Although perhaps unlikely, a future directional CNB detector might image the angular distribution of relic neutrinos. We have produced maps and power spectra of the non-linear neutrino perturbations imprinted by the large-scale structure. Our findings are in qualitative agreement with the linear theory results of \cite{hannestad09} for masses $m_\nu\leq\SI{0.1}{\eV}$, but with a much larger gravitational effect that produces an oscillatory feature in the power spectrum. This feature is related to cosmic variance in the dark matter density field. Indeed, we find that local neutrino density perturbations, in principle detectable from Earth, are anti-correlated with the projected dark matter density up to at least $\SI{250}{\Mpc}$, the largest distance constrained by the \texttt{2M++} catalogue, although for very nearby structures and fast neutrinos, we instead predict a positive correlation. The distance at which neutrinos are most sensitive to the intervening cosmic structure increases with momentum and decreases with mass, potentially enabling a kind of neutrino tomography of the large-scale structure, which would be impervious to extinction by gas and dust. Finally, we presented maps of the forecasted neutrino distribution in the local Universe, identifying neutrino clouds associated with several well-known clusters, such as Coma and Hercules. We release our simulation data to the public, which we hope will be useful for future analyses of the neutrino background.

\acknowledgments

We thank the anonymous referee for detailed and insightful comments. We are grateful to the authors of \textsc{class}, \textsc{gadget-4}, and \textsc{monofonIC} for making their codes available to the public. WE is supported by the Durham Prize Scholarship in Astroparticle Physics. We acknowledge support from the European Research Council through ERC Advanced Investigator grant, DMIDAS [GA 786910] to CSF. BL is supported by the European Research Council (ERC) through ERC starting Grant No.~716532. WE, CSF, AJ, and BL acknowledge STFC Consolidated Grants ST/P000541/1, ST/T000244/1, ST/X001075/1. SP acknowledges partial support from the European Union’s Horizon 2020 research and innovation programme under the Marie Sklodowska-Curie grant agreement No. H2020-MSCAITN-2019/860881-HIDDeN (ITN HIDDeN) and from the European Union’s Horizon Europe research and innovation programme under the Marie Sklodowska-Curie Staff Exchange grant agreement No. 101086085 -- ASYMMETRY. JJ acknowledges support by the Swedish Research Council (VR) under the project 2020-05143 -- ``Deciphering the Dynamics of Cosmic Structure''. GL acknowledges support by the ANR BIG4 project, grant ANR-16-CE23-0002 of the French Agence Nationale de la Recherche, and the grant GCEuclid from ``Centre National d'Etudes Spatiales'' (CNES). This work was supported by the Simons Collaboration on ``Learning the Universe''. This work used the DiRAC@Durham facility managed by the Institute for Computational Cosmology on behalf of the STFC DiRAC HPC Facility (www.dirac.ac.uk). The equipment was funded by BEIS capital funding via STFC capital grants ST/K00042X/1, ST/P002293/1 and ST/R002371/1, Durham University and STFC operations grant ST/R000832/1. DiRAC is part of the National e-Infrastructure.

\appendix

\section{Data availability}\label{sec:opendata}

The simulation data are available at \url{https://www.willemelbers.com/files/nubg_data/}. There are $2\times9\times6$ simulation files, corresponding to the versions with and without Milky Way, for nine posterior realizations of the initial conditions, and six different neutrino masses. For each spectator particle, we provide the phase-space density $g$, the sampled final peculiar velocity vector $\mathbf{v}_f$ at $z=0$, its backtraced initial peculiar velocity $\mathbf{v}_i$ at $z=31$, its initial and final comoving coordinates, and its statistical phase-space weight $w=(\bar{f}-g)/g$. Phase space statistics can be computed using Eq.~\eqref{eq:phase_stats}. For instance, the perturbation to the local number density is simply given by the average weight: $n/\bar{n}=1+\langle w\rangle$. We provide analysis scripts at \url{https://github.com/wullm/nubg_scripts}.

\section{Reversible simulations}\label{sec:reversible}

Running a cosmological $N$-body simulation backwards to recover the initial conditions is non-trivial (see \cite{levesque93,li10,rein17} for related ideas). In principle, leapfrog integration is time-reversible \cite{quinn97}. However, in practice, small rounding errors inevitably accumulate in the backwards direction. This is problematic if one aims to recover a low entropy initial configuration (such as two merging galaxies that are initially well separated) from a final high entropy configuration (the merged galaxy). The root of the problem is the non-associativity of standard floating point arithmetic, causing different rounding errors in backwards integrations. Furthermore, floating point errors are not necessarily reproducible in parallel programs, because of the unpredictable execution order of threads. We here briefly discuss the modifications necessary to make a cosmological code reversible, in anticipation that this may be useful for other applications.

To test the bitwise reversibility of \textsc{Gadget}-4, we periodically compute a hash of all particle data. The state of the simulation should be identical in the forwards and backwards directions at the beginning and end, respectively, of each corresponding step. Unsurprisingly, the code is not reversible by default. A first step towards achieving this is to store particle positions and velocities as integers. Implementing integer velocities is a natural step, because \textsc{Gadget}-4 already uses integer positions by default to achieve constant precision throughout the simulation domain \cite{springel20}. However, this is by no means enough to guarantee reversibility, if only because the gravitational Tree-PM algorithm still relies on floating point operations.

To guarantee reversibility, we must therefore also ensure that different threads execute their tree calculations in the same order in both directions. Furthermore, there can be no time-asymmetric decision making. For instance, we use a basic geometric tree opening criterion \cite{barnes86}, because the more adaptive opening criterion available in \textsc{Gadget}-4 depends on the particle accelerations from the previous step, which are different in the backwards direction. Similarly, the time step is usually chosen based on the maximum distance that particles can move or on the acceleration of particles in the previous time step, which again introduces an asymmetry. To address this problem without adopting a constant time step, we store a list of step sizes used in the forwards direction and feed this file back in the backwards direction. Special consideration is also needed for the neutrinos to ensure that the $\delta f$ weighting is time-reversible. Special relativistic velocities \eqref{eq:x_eq} can be absorbed in the leapfrog integration scheme \cite{elbers22b}.

The domain decomposition is another point of concern. By default, \textsc{Gadget}-4 uses floating point arithmetic for load balancing, which can lead to differences between the forwards and backwards runs. These operations are therefore modified to use integers. As a final example, recall that we inject additional `spectator' neutrinos at the start of the backwards runs. We take steps to ensure that their presence affects neither the domain decomposition of the original particles nor alters the gravity calculation. With these modifications, we exactly recover the initial conditions when running our constrained neutrino simulations backwards.



\bibliographystyle{JHEP}
\bibliography{main}

\providecommand{\href}[2]{#2}\begingroup\raggedright\begin{thebibliography}{10}

\bibitem{cobe96}
C.L.~Bennett, A.~Banday, K.M.~Gorski, G.~Hinshaw, P.~Jackson, P.~Keegstra
  et~al., \emph{{Four year COBE DMR cosmic microwave background observations:
  Maps and basic results}},
  \href{https://doi.org/10.1086/310075}{\emph{Astrophys. J. Lett.} {\bfseries
  464} (1996) L1} [\href{https://arxiv.org/abs/astro-ph/9601067}{{\ttfamily
  astro-ph/9601067}}].

\bibitem{wmap11}
{\scshape WMAP} collaboration, \emph{{Seven-Year Wilkinson Microwave Anisotropy
  Probe (WMAP) Observations: Cosmological Interpretation}},
  \href{https://doi.org/10.1088/0067-0049/192/2/18}{\emph{Astrophys. J. Suppl.}
  {\bfseries 192} (2011) 18} [\href{https://arxiv.org/abs/1001.4538}{{\ttfamily
  1001.4538}}].

\bibitem{planck18b}
{\scshape Planck} collaboration, \emph{{Planck 2018 results. I. Overview and
  the cosmological legacy of Planck}},
  \href{https://doi.org/10.1051/0004-6361/201833880}{\emph{Astron. Astrophys.}
  {\bfseries 641} (2020) A1}
  [\href{https://arxiv.org/abs/1807.06205}{{\ttfamily 1807.06205}}].

\bibitem{planck18a}
{\scshape Planck} collaboration, \emph{{Planck 2018 results. VI. Cosmological
  parameters}},
  \href{https://doi.org/10.1051/0004-6361/201833910}{\emph{Astron. Astrophys.}
  {\bfseries 641} (2020) A6}
  [\href{https://arxiv.org/abs/1807.06209}{{\ttfamily 1807.06209}}].

\bibitem{weinberg62}
S.~Weinberg, \emph{{Universal Neutrino Degeneracy}},
  \href{https://doi.org/10.1103/PhysRev.128.1457}{\emph{Phys. Rev.} {\bfseries
  128} (1962) 1457}.

\bibitem{mangano05}
G.~Mangano, G.~Miele, S.~Pastor, T.~Pinto, O.~Pisanti and P.D.~Serpico,
  \emph{{Relic neutrino decoupling including flavor oscillations}},
  \href{https://doi.org/10.1016/j.nuclphysb.2005.09.041}{\emph{Nucl. Phys. B}
  {\bfseries 729} (2005) 221}
  [\href{https://arxiv.org/abs/hep-ph/0506164}{{\ttfamily hep-ph/0506164}}].

\bibitem{desalas16}
P.F.~de~Salas and S.~Pastor, \emph{{Relic neutrino decoupling with flavour
  oscillations revisited}},
  \href{https://doi.org/10.1088/1475-7516/2016/07/051}{\emph{JCAP} {\bfseries
  07} (2016) 051} [\href{https://arxiv.org/abs/1606.06986}{{\ttfamily
  1606.06986}}].

\bibitem{fukuda98}
{\scshape Super-Kamiokande} collaboration, \emph{{Measurement of the solar
  neutrino energy spectrum using neutrino electron scattering}},
  \href{https://doi.org/10.1103/PhysRevLett.82.2430}{\emph{Phys. Rev. Lett.}
  {\bfseries 82} (1999) 2430}
  [\href{https://arxiv.org/abs/hep-ex/9812011}{{\ttfamily hep-ex/9812011}}].

\bibitem{ahmad02}
{\scshape SNO} collaboration, \emph{{Measurement of day and night neutrino
  energy spectra at SNO and constraints on neutrino mixing parameters}},
  \href{https://doi.org/10.1103/PhysRevLett.89.011302}{\emph{Phys. Rev. Lett.}
  {\bfseries 89} (2002) 011302}
  [\href{https://arxiv.org/abs/nucl-ex/0204009}{{\ttfamily nucl-ex/0204009}}].

\bibitem{esteban20}
I.~Esteban, M.C.~Gonzalez-Garcia, M.~Maltoni, T.~Schwetz and A.~Zhou,
  \emph{{The fate of hints: updated global analysis of three-flavor neutrino
  oscillations}}, \href{https://doi.org/10.1007/JHEP09(2020)178}{\emph{JHEP}
  {\bfseries 09} (2020) 178}
  [\href{https://arxiv.org/abs/2007.14792}{{\ttfamily 2007.14792}}].

\bibitem{palanque20}
N.~Palanque-Delabrouille, C.~Y\`eche, N.~Sch\"oneberg, J.~Lesgourgues,
  M.~Walther, S.~Chabanier et~al., \emph{{Hints, neutrino bounds and WDM
  constraints from SDSS DR14 Lyman-$\alpha$ and Planck full-survey data}},
  \href{https://doi.org/10.1088/1475-7516/2020/04/038}{\emph{JCAP} {\bfseries
  04} (2020) 038} [\href{https://arxiv.org/abs/1911.09073}{{\ttfamily
  1911.09073}}].

\bibitem{choudhury20}
S.~Roy~Choudhury and S.~Hannestad, \emph{{Updated results on neutrino mass and
  mass hierarchy from cosmology with Planck 2018 likelihoods}},
  \href{https://doi.org/10.1088/1475-7516/2020/07/037}{\emph{JCAP} {\bfseries
  07} (2020) 037} [\href{https://arxiv.org/abs/1907.12598}{{\ttfamily
  1907.12598}}].

\bibitem{DES21}
{\scshape DES} collaboration, \emph{{Dark Energy Survey Year 3 results:
  Cosmological constraints from galaxy clustering and weak lensing}},
  \href{https://doi.org/10.1103/PhysRevD.105.023520}{\emph{Phys. Rev. D}
  {\bfseries 105} (2022) 023520}
  [\href{https://arxiv.org/abs/2105.13549}{{\ttfamily 2105.13549}}].

\bibitem{divalentino21}
E.~Di~Valentino, S.~Gariazzo and O.~Mena, \emph{{Most constraining cosmological
  neutrino mass bounds}},
  \href{https://doi.org/10.1103/PhysRevD.104.083504}{\emph{Phys. Rev. D}
  {\bfseries 104} (2021) 083504}
  [\href{https://arxiv.org/abs/2106.15267}{{\ttfamily 2106.15267}}].

\bibitem{katrin22}
{\scshape KATRIN} collaboration, \emph{{New Constraint on the Local Relic
  Neutrino Background Overdensity with the First KATRIN Data Runs}},
  \href{https://doi.org/10.1103/PhysRevLett.129.011806}{\emph{Phys. Rev. Lett.}
  {\bfseries 129} (2022) 011806}
  [\href{https://arxiv.org/abs/2202.04587}{{\ttfamily 2202.04587}}].

\bibitem{ptolemy18}
{\scshape PTOLEMY} collaboration, \emph{{PTOLEMY: A Proposal for Thermal Relic
  Detection of Massive Neutrinos and Directional Detection of MeV Dark
  Matter}},  \href{https://arxiv.org/abs/1808.01892}{{\ttfamily 1808.01892}}.

\bibitem{betti19}
{\scshape PTOLEMY} collaboration, \emph{{Neutrino physics with the PTOLEMY
  project: active neutrino properties and the light sterile case}},
  \href{https://doi.org/10.1088/1475-7516/2019/07/047}{\emph{JCAP} {\bfseries
  07} (2019) 047} [\href{https://arxiv.org/abs/1902.05508}{{\ttfamily
  1902.05508}}].

\bibitem{ptolemy22}
{\scshape PTOLEMY} collaboration, \emph{{Heisenberg\textquoteright{}s
  uncertainty principle in the PTOLEMY project: A theory update}},
  \href{https://doi.org/10.1103/PhysRevD.106.053002}{\emph{Phys. Rev. D}
  {\bfseries 106} (2022) 053002}
  [\href{https://arxiv.org/abs/2203.11228}{{\ttfamily 2203.11228}}].

\bibitem{cocco07}
A.G.~Cocco, G.~Mangano and M.~Messina, \emph{{Probing low energy neutrino
  backgrounds with neutrino capture on beta decaying nuclei}},
  \href{https://doi.org/10.1088/1475-7516/2007/06/015}{\emph{JCAP} {\bfseries
  06} (2007) 015} [\href{https://arxiv.org/abs/hep-ph/0703075}{{\ttfamily
  hep-ph/0703075}}].

\bibitem{cheipesh21}
Y.~Cheipesh, V.~Cheianov and A.~Boyarsky, \emph{{Navigating the pitfalls of
  relic neutrino detection}},
  \href{https://doi.org/10.1103/PhysRevD.104.116004}{\emph{Phys. Rev. D}
  {\bfseries 104} (2021) 116004}
  [\href{https://arxiv.org/abs/2101.10069}{{\ttfamily 2101.10069}}].

\bibitem{opher74}
R.~Opher, \emph{{Coherent scattering of cosmic neutrinos}}, {\emph{Astron.
  Astrophys.} {\bfseries 37} (1974) 135}.

\bibitem{stodolsky74}
L.~Stodolsky, \emph{{Speculations on Detection of the Neutrino Sea}},
  \href{https://doi.org/10.1103/PhysRevLett.34.110}{\emph{Phys. Rev. Lett.}
  {\bfseries 34} (1975) 110}.

\bibitem{domcke17}
V.~Domcke and M.~Spinrath, \emph{{Detection prospects for the Cosmic Neutrino
  Background using laser interferometers}},
  \href{https://doi.org/10.1088/1475-7516/2017/06/055}{\emph{JCAP} {\bfseries
  06} (2017) 055} [\href{https://arxiv.org/abs/1703.08629}{{\ttfamily
  1703.08629}}].

\bibitem{shergold21}
J.D.~Shergold, \emph{{Updated detection prospects for relic neutrinos using
  coherent scattering}},
  \href{https://doi.org/10.1088/1475-7516/2021/11/052}{\emph{JCAP} {\bfseries
  11} (2021) 052} [\href{https://arxiv.org/abs/2109.07482}{{\ttfamily
  2109.07482}}].

\bibitem{eberle04}
B.~Eberle, A.~Ringwald, L.~Song and T.J.~Weiler, \emph{{Relic neutrino
  absorption spectroscopy}},
  \href{https://doi.org/10.1103/PhysRevD.70.023007}{\emph{Phys. Rev. D}
  {\bfseries 70} (2004) 023007}
  [\href{https://arxiv.org/abs/hep-ph/0401203}{{\ttfamily hep-ph/0401203}}].

\bibitem{brdar22}
V.~Brdar, P.S.B.~Dev, R.~Plestid and A.~Soni, \emph{{A new probe of relic
  neutrino clustering using cosmogenic neutrinos}},
  \href{https://doi.org/10.1016/j.physletb.2022.137358}{\emph{Phys. Lett. B}
  {\bfseries 833} (2022) 137358}
  [\href{https://arxiv.org/abs/2207.02860}{{\ttfamily 2207.02860}}].

\bibitem{yoshimura14}
M.~Yoshimura, N.~Sasao and M.~Tanaka, \emph{{Experimental method of detecting
  relic neutrino by atomic de-excitation}},
  \href{https://doi.org/10.1103/PhysRevD.91.063516}{\emph{Phys. Rev. D}
  {\bfseries 91} (2015) 063516}
  [\href{https://arxiv.org/abs/1409.3648}{{\ttfamily 1409.3648}}].

\bibitem{bauer21}
M.~Bauer and J.D.~Shergold, \emph{{Relic neutrinos at accelerator
  experiments}}, \href{https://doi.org/10.1103/PhysRevD.104.083039}{\emph{Phys.
  Rev. D} {\bfseries 104} (2021) 083039}
  [\href{https://arxiv.org/abs/2104.12784}{{\ttfamily 2104.12784}}].

\bibitem{bauer22}
M.~Bauer and J.D.~Shergold, \emph{{Limits on the cosmic neutrino background}},
  \href{https://doi.org/10.1088/1475-7516/2023/01/003}{\emph{JCAP} {\bfseries
  01} (2023) 003} [\href{https://arxiv.org/abs/2207.12413}{{\ttfamily
  2207.12413}}].

\bibitem{hu95}
W.~Hu, D.~Scott, N.~Sugiyama and M.J.~White, \emph{{The Effect of physical
  assumptions on the calculation of microwave background anisotropies}},
  \href{https://doi.org/10.1103/PhysRevD.52.5498}{\emph{Phys. Rev. D}
  {\bfseries 52} (1995) 5498}
  [\href{https://arxiv.org/abs/astro-ph/9505043}{{\ttfamily
  astro-ph/9505043}}].

\bibitem{michney06}
R.J.~Michney and R.R.~Caldwell, \emph{{Anisotropy of the Cosmic Neutrino
  Background}},
  \href{https://doi.org/10.1088/1475-7516/2007/01/014}{\emph{JCAP} {\bfseries
  01} (2007) 014} [\href{https://arxiv.org/abs/astro-ph/0608303}{{\ttfamily
  astro-ph/0608303}}].

\bibitem{hannestad09}
S.~Hannestad and J.~Brandbyge, \emph{{The Cosmic Neutrino Background Anisotropy
  - Linear Theory}},
  \href{https://doi.org/10.1088/1475-7516/2010/03/020}{\emph{JCAP} {\bfseries
  03} (2010) 020} [\href{https://arxiv.org/abs/0910.4578}{{\ttfamily
  0910.4578}}].

\bibitem{lin19}
J.Y.-Y.~Lin and G.~Holder, \emph{{Gravitational Lensing of the Cosmic Neutrino
  Background}},
  \href{https://doi.org/10.1088/1475-7516/2020/04/054}{\emph{JCAP} {\bfseries
  04} (2020) 054} [\href{https://arxiv.org/abs/1910.03550}{{\ttfamily
  1910.03550}}].

\bibitem{tully21}
C.G.~{Tully} and G.~{Zhang}, \emph{{Multi-messenger astrophysics with the
  cosmic neutrino background}},
  \href{https://doi.org/10.1088/1475-7516/2021/06/053}{\emph{JCAP} {\bfseries
  2021} (2021) 053} [\href{https://arxiv.org/abs/2103.01274}{{\ttfamily
  2103.01274}}].

\bibitem{lisanti14}
M.~{Lisanti}, B.R.~{Safdi} and C.G.~{Tully}, \emph{{Measuring anisotropies in
  the cosmic neutrino background}},
  \href{https://doi.org/10.1103/PhysRevD.90.073006}{\emph{PRD} {\bfseries 90}
  (2014) 073006} [\href{https://arxiv.org/abs/1407.0393}{{\ttfamily
  1407.0393}}].

\bibitem{safdi14}
B.R.~Safdi, M.~Lisanti, J.~Spitz and J.A.~Formaggio, \emph{{Annual Modulation
  of Cosmic Relic Neutrinos}},
  \href{https://doi.org/10.1103/PhysRevD.90.043001}{\emph{Phys. Rev. D}
  {\bfseries 90} (2014) 043001}
  [\href{https://arxiv.org/abs/1404.0680}{{\ttfamily 1404.0680}}].

\bibitem{akhmedov19}
E.~Akhmedov, \emph{{Relic neutrino detection through angular correlations in
  inverse $\beta$-decay}},
  \href{https://doi.org/10.1088/1475-7516/2019/09/031}{\emph{JCAP} {\bfseries
  09} (2019) 031} [\href{https://arxiv.org/abs/1905.10207}{{\ttfamily
  1905.10207}}].

\bibitem{baushev20}
A.N.~Baushev, \emph{{The Relic Neutrino Composition as Seen from Earth}},
  \href{https://doi.org/10.1134/S1063772920120021}{\emph{Astron. Rep.}
  {\bfseries 64} (2020) 1005}
  [\href{https://arxiv.org/abs/2101.11405}{{\ttfamily 2101.11405}}].

\bibitem{baym21}
G.~Baym and J.-C.~Peng, \emph{{Evolution of primordial neutrino helicities in
  cosmic gravitational inhomogeneities}},
  \href{https://doi.org/10.1103/PhysRevD.103.123019}{\emph{Phys. Rev. D}
  {\bfseries 103} (2021) 123019}
  [\href{https://arxiv.org/abs/2103.11209}{{\ttfamily 2103.11209}}].

\bibitem{hernandez_molinero22}
B.~Hernandez-Molinero, R.~Jimenez and C.~Pena-Garay, \emph{{Distinguishing
  Dirac vs. Majorana neutrinos: a cosmological probe}},
  \href{https://doi.org/10.1088/1475-7516/2022/08/038}{\emph{JCAP} {\bfseries
  08} (2022) 038} [\href{https://arxiv.org/abs/2205.00808}{{\ttfamily
  2205.00808}}].

\bibitem{peng22}
J.-C.~Peng and G.~Baym, \emph{{Implication of Helicity Modifications of
  Primordial Neutrinos on Their Detection}},
  \href{https://doi.org/10.7566/JPSCP.37.020704}{\emph{JPS Conf. Proc.}
  {\bfseries 37} (2022) 020704}.

\bibitem{hernandez_molinero23}
B.~Hern\'andez-Molinero, C.~Carbone, R.~Jimenez and C.~Pe\~na Garay,
  \emph{{Cosmic Background Neutrinos Deflected by Gravity: DEMNUni Simulation
  Analysis}},  \href{https://arxiv.org/abs/2301.12430}{{\ttfamily 2301.12430}}.

\bibitem{ringwald04}
A.~Ringwald and Y.Y.Y.~Wong, \emph{{Gravitational clustering of relic neutrinos
  and implications for their detection}},
  \href{https://doi.org/10.1088/1475-7516/2004/12/005}{\emph{JCAP} {\bfseries
  12} (2004) 005} [\href{https://arxiv.org/abs/hep-ph/0408241}{{\ttfamily
  hep-ph/0408241}}].

\bibitem{desalas17}
P.F.~de~Salas, S.~Gariazzo, J.~Lesgourgues and S.~Pastor, \emph{{Calculation of
  the local density of relic neutrinos}},
  \href{https://doi.org/10.1088/1475-7516/2017/09/034}{\emph{JCAP} {\bfseries
  09} (2017) 034} [\href{https://arxiv.org/abs/1706.09850}{{\ttfamily
  1706.09850}}].

\bibitem{zhang17}
J.~Zhang and X.~Zhang, \emph{{Gravitational clustering of cosmic relic
  neutrinos in the Milky Way}},
  \href{https://doi.org/10.1038/s41467-018-04264-y}{\emph{Nature Commun.}
  {\bfseries 9} (2018) 1833}
  [\href{https://arxiv.org/abs/1712.01153}{{\ttfamily 1712.01153}}].

\bibitem{mertsch19}
P.~Mertsch, G.~Parimbelli, P.F.~de~Salas, S.~Gariazzo, J.~Lesgourgues and
  S.~Pastor, \emph{{Neutrino clustering in the Milky Way and beyond}},
  \href{https://doi.org/10.1088/1475-7516/2020/01/015}{\emph{JCAP} {\bfseries
  01} (2020) 015} [\href{https://arxiv.org/abs/1910.13388}{{\ttfamily
  1910.13388}}].

\bibitem{holm23}
E.~{Brinch Holm}, I.M.~{Oldengott} and S.~{Zentarra}, \emph{{Local clustering
  of relic neutrinos with kinetic field theory}},
  \href{https://doi.org/10.48550/arXiv.2305.13379}{\emph{arXiv e-prints} (2023)
  arXiv:2305.13379} [\href{https://arxiv.org/abs/2305.13379}{{\ttfamily
  2305.13379}}].

\bibitem{zimmer23}
F.~Zimmer, C.A.~Correa and S.~Ando, \emph{{Influence of local structure on
  relic neutrino abundances and anisotropies}},
  \href{https://arxiv.org/abs/2306.16444}{{\ttfamily 2306.16444}}.

\bibitem{elbers21}
W.~Elbers, C.S.~Frenk, A.~Jenkins, B.~Li and S.~Pascoli, \emph{{An optimal
  non-linear method for simulating relic neutrinos}},
  \href{https://doi.org/10.1093/mnras/stab2260}{\emph{Mon. Not. Roy. Astron.
  Soc.} {\bfseries 507} (2021) 2614}
  [\href{https://arxiv.org/abs/2010.07321}{{\ttfamily 2010.07321}}].

\bibitem{lavaux11}
G.~Lavaux and M.J.~Hudson, \emph{{The 2M++ galaxy redshift catalogue}},
  \href{https://doi.org/10.1111/j.1365-2966.2011.19233.x}{\emph{Mon. Not. Roy.
  Astron. Soc.} {\bfseries 416} (2011) 2840}
  [\href{https://arxiv.org/abs/1105.6107}{{\ttfamily 1105.6107}}].

\bibitem{cautun19}
M.~Cautun, A.~Benitez-Llambay, A.J.~Deason, C.S.~Frenk, A.~Fattahi,
  F.A.~G\'omez et~al., \emph{{The Milky Way total mass profile as inferred from
  Gaia DR2}}, \href{https://doi.org/10.1093/mnras/staa1017}{\emph{Mon. Not.
  Roy. Astron. Soc.} {\bfseries 494} (2020) 4291}
  [\href{https://arxiv.org/abs/1911.04557}{{\ttfamily 1911.04557}}].

\bibitem{yepes13}
G.~Yepes, S.~Gottl\"ober and Y.~Hoffman, \emph{{Dark Matter in the Local
  Universe}}, \href{https://doi.org/10.1016/j.newar.2013.11.001}{\emph{New
  Astron. Rev.} {\bfseries 58} (2014) 1}
  [\href{https://arxiv.org/abs/1312.0105}{{\ttfamily 1312.0105}}].

\bibitem{wang14}
H.~Wang, H.J.~Mo, X.~Yang, Y.P.~Jing and W.P.~Lin, \emph{{ELUCID - Exploring
  the Local Universe with reConstructed Initial Density field I: Hamiltonian
  Markov Chain Monte Carlo Method with Particle Mesh Dynamics}},
  \href{https://doi.org/10.1088/0004-637X/794/1/94}{\emph{Astrophys. J.}
  {\bfseries 794} (2014) 94} [\href{https://arxiv.org/abs/1407.3451}{{\ttfamily
  1407.3451}}].

\bibitem{libeskind20}
N.I.~Libeskind, E.~Carlesi, R.J.~Grand, A.~Khalatyan, A.~Knebe, R.~Pakmor
  et~al., \emph{The hestia project: simulations of the local group},
  {\emph{MNRAS} {\bfseries 498} (2020) 2968}.

\bibitem{sorce21}
J.G.~Sorce, Y.~Dubois, J.~Blaizot, S.L.~McGee, G.~Yepes and A.~Knebe, \emph{{I
  \textendash{} A hydrodynamical clone of the Virgo cluster of galaxies to
  confirm observationally driven formation scenarios}},
  \href{https://doi.org/10.1093/mnras/stab1021}{\emph{Mon. Not. Roy. Astron.
  Soc.} {\bfseries 504} (2021) 2998}
  [\href{https://arxiv.org/abs/2104.13389}{{\ttfamily 2104.13389}}].

\bibitem{mcalpine22}
S.~McAlpine, J.C.~Helly, M.~Schaller, T.~Sawala, G.~Lavaux, J.~Jasche et~al.,
  \emph{{SIBELIUS-DARK: a galaxy catalogue of the local volume from a
  constrained realization simulation}},
  \href{https://doi.org/10.1093/mnras/stac295}{\emph{Mon. Not. Roy. Astron.
  Soc.} {\bfseries 512} (2022) 5823}
  [\href{https://arxiv.org/abs/2202.04099}{{\ttfamily 2202.04099}}].

\bibitem{oei22}
M.S.S.L.~Oei, R.J.~van Weeren, F.~Vazza, F.~Leclercq, A.~Gopinath and
  H.J.A.~R\"ottgering, \emph{{Filamentary baryons and where to find them - A
  forecast of synchrotron radiation from merger and accretion shocks in the
  local Cosmic Web}},
  \href{https://doi.org/10.1051/0004-6361/202140364}{\emph{Astron. Astrophys.}
  {\bfseries 662} (2022) A87}
  [\href{https://arxiv.org/abs/2203.05365}{{\ttfamily 2203.05365}}].

\bibitem{jasche12}
J.~Jasche and B.D.~Wandelt, \emph{{Bayesian physical reconstruction of initial
  conditions from large scale structure surveys}},
  \href{https://doi.org/10.1093/mnras/stt449}{\emph{Mon. Not. Roy. Astron.
  Soc.} {\bfseries 432} (2013) 894}
  [\href{https://arxiv.org/abs/1203.3639}{{\ttfamily 1203.3639}}].

\bibitem{jasche19}
J.~Jasche and G.~Lavaux, \emph{{Physical Bayesian modelling of the non-linear
  matter distribution: new insights into the Nearby Universe}},
  \href{https://doi.org/10.1051/0004-6361/201833710}{\emph{Astron. Astrophys.}
  {\bfseries 625} (2019) A64}
  [\href{https://arxiv.org/abs/1806.11117}{{\ttfamily 1806.11117}}].

\bibitem{stopyra23}
S.~{Stopyra}, H.V.~{Peiris}, A.~{Pontzen}, J.~{Jasche} and G.~{Lavaux},
  \emph{{Towards Accurate Field-Level Inference of Massive Cosmic Structures}},
  \href{https://doi.org/10.48550/arXiv.2304.09193}{\emph{arXiv e-prints} (2023)
  arXiv:2304.09193} [\href{https://arxiv.org/abs/2304.09193}{{\ttfamily
  2304.09193}}].

\bibitem{tassev13}
S.~Tassev, M.~Zaldarriaga and D.~Eisenstein, \emph{{Solving Large Scale
  Structure in Ten Easy Steps with COLA}},
  \href{https://doi.org/10.1088/1475-7516/2013/06/036}{\emph{JCAP} {\bfseries
  06} (2013) 036} [\href{https://arxiv.org/abs/1301.0322}{{\ttfamily
  1301.0322}}].

\bibitem{bartlett21}
D.J.~{Bartlett}, H.~{Desmond} and P.G.~{Ferreira}, \emph{{Constraints on
  Galileons from the positions of supermassive black holes}},
  \href{https://doi.org/10.1103/PhysRevD.103.023523}{\emph{PRD} {\bfseries 103}
  (2021) 023523} [\href{https://arxiv.org/abs/2010.05811}{{\ttfamily
  2010.05811}}].

\bibitem{bartlett22}
D.J.~{Bartlett}, A.~{Kosti{\'c}}, H.~{Desmond}, J.~{Jasche} and G.~{Lavaux},
  \emph{{Constraints on dark matter annihilation and decay from the large-scale
  structure of the nearby Universe}},
  \href{https://doi.org/10.1103/PhysRevD.106.103526}{\emph{PRD} {\bfseries 106}
  (2022) 103526} [\href{https://arxiv.org/abs/2205.12916}{{\ttfamily
  2205.12916}}].

\bibitem{desmond22}
H.~{Desmond}, M.L.~{Hutt}, J.~{Devriendt} and A.~{Slyz}, \emph{{Catalogues of
  voids as antihaloes in the local Universe}},
  \href{https://doi.org/10.1093/mnrasl/slac008}{\emph{MNRAS} {\bfseries 511}
  (2022) L45} [\href{https://arxiv.org/abs/2109.09439}{{\ttfamily
  2109.09439}}].

\bibitem{hutt22}
M.L.~{Hutt}, H.~{Desmond}, J.~{Devriendt} and A.~{Slyz}, \emph{{The effect of
  local Universe constraints on halo abundance and clustering}},
  \href{https://doi.org/10.1093/mnras/stac2407}{\emph{MNRAS} {\bfseries 516}
  (2022) 3592} [\href{https://arxiv.org/abs/2203.14724}{{\ttfamily
  2203.14724}}].

\bibitem{michaux20}
M.~Michaux, O.~Hahn, C.~Rampf and R.E.~Angulo, \emph{{Accurate initial
  conditions for cosmological N-body simulations: Minimizing truncation and
  discreteness errors}},
  \href{https://doi.org/10.1093/mnras/staa3149}{\emph{Mon. Not. Roy. Astron.
  Soc.} {\bfseries 500} (2020) 663}
  [\href{https://arxiv.org/abs/2008.09588}{{\ttfamily 2008.09588}}].

\bibitem{hahn20}
O.~Hahn, M.~Michaux, C.~Rampf, C.~Uhlemann and R.E.~Angulo,
  \emph{{MUSIC2-monofonIC: 3LPT initial condition generator}},
  {\emph{Astrophysics Source Code Library} (2020) ascl}.

\bibitem{elbers22}
W.~Elbers, C.S.~Frenk, A.~Jenkins, B.~Li and S.~Pascoli, \emph{{Higher order
  initial conditions with massive neutrinos}},
  \href{https://doi.org/10.1093/mnras/stac2365}{\emph{Mon. Not. Roy. Astron.
  Soc.} {\bfseries 516} (2022) 3821}
  [\href{https://arxiv.org/abs/2202.00670}{{\ttfamily 2202.00670}}].

\bibitem{elbers22b}
W.~Elbers, \emph{{Geodesic motion and phase-space evolution of massive
  neutrinos}}, \href{https://doi.org/10.1088/1475-7516/2022/11/058}{\emph{JCAP}
  {\bfseries 11} (2022) 058}
  [\href{https://arxiv.org/abs/2207.14256}{{\ttfamily 2207.14256}}].

\bibitem{lesgourgues11}
J.~Lesgourgues, \emph{{The Cosmic Linear Anisotropy Solving System (CLASS) I:
  Overview}},  \href{https://arxiv.org/abs/1104.2932}{{\ttfamily 1104.2932}}.

\bibitem{lesgourgues11b}
J.~Lesgourgues and T.~Tram, \emph{{The Cosmic Linear Anisotropy Solving System
  (CLASS) IV: efficient implementation of non-cold relics}},
  \href{https://doi.org/10.1088/1475-7516/2011/09/032}{\emph{JCAP} {\bfseries
  09} (2011) 032} [\href{https://arxiv.org/abs/1104.2935}{{\ttfamily
  1104.2935}}].

\bibitem{springel20}
V.~Springel, R.~Pakmor, O.~Zier and M.~Reinecke, \emph{{Simulating cosmic
  structure formation with the gadget-4 code}},
  \href{https://doi.org/10.1093/mnras/stab1855}{\emph{Mon. Not. Roy. Astron.
  Soc.} {\bfseries 506} (2021) 2871}
  [\href{https://arxiv.org/abs/2010.03567}{{\ttfamily 2010.03567}}].

\bibitem{navarro96}
J.F.~Navarro, C.S.~Frenk and S.D.M.~White, \emph{{A Universal density profile
  from hierarchical clustering}},
  \href{https://doi.org/10.1086/304888}{\emph{Astrophys. J.} {\bfseries 490}
  (1997) 493} [\href{https://arxiv.org/abs/astro-ph/9611107}{{\ttfamily
  astro-ph/9611107}}].

\bibitem{tully07}
R.B.~Tully, E.J.~Shaya, I.D.~Karachentsev, H.M.~Courtois, D.D.~Kocevski,
  L.~Rizzi et~al., \emph{{Our Peculiar Motion Away from the Local Void}},
  \href{https://doi.org/10.1086/527428}{\emph{Astrophys. J.} {\bfseries 676}
  (2008) 184} [\href{https://arxiv.org/abs/0705.4139}{{\ttfamily 0705.4139}}].

\bibitem{tully16}
R.B.~Tully, H.M.~Courtois and J.G.~Sorce, \emph{{Cosmicflows-3}},
  \href{https://doi.org/10.3847/0004-6256/152/2/50}{\emph{Astron. J.}
  {\bfseries 152} (2016) 50}
  [\href{https://arxiv.org/abs/1605.01765}{{\ttfamily 1605.01765}}].

\bibitem{ichiki11}
K.~Ichiki and M.~Takada, \emph{{The impact of massive neutrinos on the
  abundance of massive clusters}},
  \href{https://doi.org/10.1103/PhysRevD.85.063521}{\emph{Phys. Rev. D}
  {\bfseries 85} (2012) 063521}
  [\href{https://arxiv.org/abs/1108.4688}{{\ttfamily 1108.4688}}].

\bibitem{castorina13}
E.~Castorina, E.~Sefusatti, R.K.~Sheth, F.~Villaescusa-Navarro and M.~Viel,
  \emph{{Cosmology with massive neutrinos II: on the universality of the halo
  mass function and bias}},
  \href{https://doi.org/10.1088/1475-7516/2014/02/049}{\emph{JCAP} {\bfseries
  02} (2014) 049} [\href{https://arxiv.org/abs/1311.1212}{{\ttfamily
  1311.1212}}].

\bibitem{massara14}
E.~Massara, F.~Villaescusa-Navarro and M.~Viel, \emph{{The halo model in a
  massive neutrino cosmology}},
  \href{https://doi.org/10.1088/1475-7516/2014/12/053}{\emph{JCAP} {\bfseries
  12} (2014) 053} [\href{https://arxiv.org/abs/1410.6813}{{\ttfamily
  1410.6813}}].

\bibitem{archidiacono20}
M.~{Archidiacono}, S.~{Hannestad} and J.~{Lesgourgues}, \emph{{What will it
  take to measure individual neutrino mass states using cosmology?}},
  \href{https://doi.org/10.1088/1475-7516/2020/09/021}{\emph{JCAP} {\bfseries
  2020} (2020) 021} [\href{https://arxiv.org/abs/2003.03354}{{\ttfamily
  2003.03354}}].

\bibitem{pontecorvo68}
B.~{Pontecorvo}, \emph{{Neutrino Experiments and the Problem of Conservation of
  Leptonic Charge}}, {\emph{Soviet Journal of Experimental and Theoretical
  Physics} {\bfseries 26} (1968)
  \href{https://ui.adsabs.harvard.edu/abs/1968JETP...26..984P}{984}}.

\bibitem{maki62}
Z.~{Maki}, M.~{Nakagawa} and S.~{Sakata}, \emph{{Remarks on the Unified Model
  of Elementary Particles}},
  \href{https://doi.org/10.1143/PTP.28.870}{\emph{Progress of Theoretical
  Physics} {\bfseries 28} (1962) 870}.

\bibitem{long14}
A.J.~Long, C.~Lunardini and E.~Sabancilar, \emph{{Detecting non-relativistic
  cosmic neutrinos by capture on tritium: phenomenology and physics
  potential}}, \href{https://doi.org/10.1088/1475-7516/2014/08/038}{\emph{JCAP}
  {\bfseries 08} (2014) 038} [\href{https://arxiv.org/abs/1405.7654}{{\ttfamily
  1405.7654}}].

\bibitem{roldan16}
O.~{Roldan}, A.~{Notari} and M.~{Quartin}, \emph{{Interpreting the CMB
  aberration and Doppler measurements: boost or intrinsic dipole?}},
  \href{https://doi.org/10.1088/1475-7516/2016/06/026}{\emph{JCAP} {\bfseries
  2016} (2016) 026} [\href{https://arxiv.org/abs/1603.02664}{{\ttfamily
  1603.02664}}].

\bibitem{meerburg17}
P.D.~{Meerburg}, J.~{Meyers} and A.~{van Engelen}, \emph{{Reconstructing the
  primary CMB dipole}},
  \href{https://doi.org/10.1103/PhysRevD.96.083519}{\emph{PRD} {\bfseries 96}
  (2017) 083519} [\href{https://arxiv.org/abs/1704.00718}{{\ttfamily
  1704.00718}}].

\bibitem{silveira20}
P.~{da Silveira Ferreira} and M.~{Quartin}, \emph{{First constraints on the
  intrinsic CMB dipole and our velocity with Doppler and aberration}},
  \href{https://doi.org/10.48550/arXiv.2011.08385}{\emph{arXiv e-prints} (2020)
  arXiv:2011.08385} [\href{https://arxiv.org/abs/2011.08385}{{\ttfamily
  2011.08385}}].

\bibitem{lewis06}
A.~Lewis and A.~Challinor, \emph{{Weak gravitational lensing of the CMB}},
  \href{https://doi.org/10.1016/j.physrep.2006.03.002}{\emph{Phys. Rept.}
  {\bfseries 429} (2006) 1}
  [\href{https://arxiv.org/abs/astro-ph/0601594}{{\ttfamily
  astro-ph/0601594}}].

\bibitem{meusinger20}
H.~{Meusinger}, C.~{Rudolf}, B.~{Stecklum}, M.~{Hoeft}, R.~{Mauersberger} and
  D.~{Apai}, \emph{{The galaxy population within the virial radius of the
  Perseus cluster}},
  \href{https://doi.org/10.1051/0004-6361/202037574}{\emph{Astron. Astrophys.}
  {\bfseries 640} (2020) A30}
  [\href{https://arxiv.org/abs/2103.11209}{{\ttfamily 2103.11209}}].

\bibitem{lyndenbell88}
D.~Lynden-Bell, S.M.~Faber, D.~Burstein, R.L.~Davies, A.~Dressler,
  R.J.~Terlevich et~al., \emph{{Spectroscopy and photometry of elliptical
  galaxies. V - Galaxy streaming toward the new supergalactic center}},
  \href{https://doi.org/10.1086/166066}{\emph{Astrophys. J.} {\bfseries 326}
  (1988) 19}.

\bibitem{hoffman17}
Y.~Hoffman, D.~Pomarede, R.~Brent~Tully and H.~Courtois, \emph{{The Dipole
  Repeller}},  \href{https://arxiv.org/abs/1702.02483}{{\ttfamily 1702.02483}}.

\bibitem{levesque93}
D.~Levesque and L.~Verlet, \emph{Molecular dynamics and time reversibility},
  {\emph{Journal of Statistical Physics} {\bfseries 72} (1993) 519}.

\bibitem{li10}
B.~Li, L.J.~King, G.-B.~Zhao and H.~Zhao, \emph{{A Semi-analytic Ray-tracing
  Algorithm for Weak Lensing}},
  \href{https://doi.org/10.1111/j.1365-2966.2011.18754.x}{\emph{Mon. Not. Roy.
  Astron. Soc.} {\bfseries 415} (2011) 881}
  [\href{https://arxiv.org/abs/1012.1625}{{\ttfamily 1012.1625}}].

\bibitem{rein17}
H.~Rein and D.~Tamayo, \emph{{JANUS: A bit-wise reversible integrator for
  N-body dynamics}}, \href{https://doi.org/10.1093/mnras/stx2479}{\emph{Mon.
  Not. Roy. Astron. Soc.} {\bfseries 473} (2018) 3351}
  [\href{https://arxiv.org/abs/1704.07715}{{\ttfamily 1704.07715}}].

\bibitem{quinn97}
T.R.~Quinn, N.~Katz, J.~Stadel and G.~Lake, \emph{{Time stepping N body
  simulations}},  \href{https://arxiv.org/abs/astro-ph/9710043}{{\ttfamily
  astro-ph/9710043}}.

\bibitem{barnes86}
J.~Barnes and P.~Hut, \emph{{A Hierarchical O(N log N) Force Calculation
  Algorithm}}, \href{https://doi.org/10.1038/324446a0}{\emph{Nature} {\bfseries
  324} (1986) 446}.

\end{thebibliography}\endgroup

\end{document}